\documentclass[aps,prl,preprint,superscriptaddress]{revtex4-1}

\usepackage{amsmath}
\usepackage{amssymb}
\usepackage{multibbl}
\usepackage{graphicx}
\usepackage[hidelinks]{hyperref}
\hypersetup{
			colorlinks = true,
			linkcolor = blue,
			citecolor = blue,
}

\newbibliography{josephson_resub}
\newbibliography{supplementary_resub}

\begin{document}

\nocite{josephson_resub}{*}
\nocite{supplementary_resub}{*}


\title{Scanning Josephson spectroscopy on the atomic scale}


\author{Mallika T. Randeria}
\thanks{These authors contributed equally}
\author{Benjamin E. Feldman}
\thanks{These authors contributed equally}
\author{Ilya K. Drozdov}
\altaffiliation[Present address: ]{Brookhaven National Lab, Upton NY 11973, USA}

\author{Ali Yazdani}
\email[Email: ]{yazdani@princeton.edu}
\affiliation{Joseph Henry Laboratories in Physics, Department of Physics, Princeton University, Princeton, NJ 08544, USA}


\date{\today}

\begin{abstract}
The Josephson effect provides a direct method to probe the strength of the pairing interaction in superconductors. By measuring the phase fluctuating Josephson current between a superconducting tip of a scanning tunneling microscope (STM) and a BCS superconductor with isolated magnetic adatoms on its surface, we demonstrate that the spatial variation of the pairing order parameter can be characterized on the atomic scale. This system provides an example where the local pairing potential suppression is not directly reflected in the spectra measured via quasipartcile tunneling. Spectroscopy with such superconducting tips also show signatures of previously unexplored Andreev processes through individual impurity-bound Shiba states. The atomic resolution achieved here establishes scanning Josephson spectroscopy as a promising technique for the study of novel superconducting phases.
\end{abstract}


\maketitle


A number of novel superconducting states of matter such as those appearing in disordered superconductors, heavy fermion materials, and high-T$_{\text c}$ superconductors have been predicted to have pairing order parameters that are spatially modulated on atomic length scales. These short range spatial modulations can occur due to different mechanisms such as the inhomogeneous material properties in disordered superconductors \cite{sacepe2011,bouadim2011,gomes2007,pasupathy2008}, a momentum dependent pairing interaction, such as the Fulde-Ferrell-Larkin-Ovchinnkov (FFLO) state proposed for heavy fermion materials \cite{fulde1964,larkin1965,matsuda2007}, or the interplay between different forms of electronic ordering in the pair density waves proposed for high-T$_{\text c}$ cuprates \cite{chen2004,berg2009,lee2014}.  Although spectroscopic mapping with a scanning tunneling microscope (STM) can provide evidence for variations in the local density of states (LDOS) through quasi-particle tunneling, such measurements probe the superconducting order parameter only indirectly. If the Josephson effect can be measured and mapped on the atomic scale, then it would allow for direct characterization of the local pairing order parameter and high-resolution studies of novel superconducting phases \cite{PhysRevB.64.212506}. 

This goal has motivated previous efforts in the use of superconducting tips in STM \cite{rodrigo2004} and has led to the local observation of thermal phase fluctuating Josephson supercurrent close to the point contact regime \cite{PhysRevLett.87.097004, rodrigoEuroPHys, levy2013}. Subsequent measurements have mapped the Josephson effect on the nanometer scale, applying this technique to vortices \cite{proslier2006,PhysRevB.78.140507} and high-T$_{\text c}$ cuprates \cite{PhysRevB.80.144506,davis2015}. A major challenge in improving the resolution of these experiments has been satisfying the competing requirements of a high junction impedance necessary for imaging and a low junction impedance allowing for the strong tip-sample coupling necessary to observe the Josephson effect despite thermal fluctuations. Extending the Josephson STM measurements to millikelvin temperatures allows for mapping of the Cooper pair current at junction resistances that are compatible with atomic resolution imaging. 

In this letter, we use scanning Josephson spectroscopy to probe variations of the superconducting order parameter on the scale of a single atom. We map the strength of the phase fluctuating Josephson current between a superconducting Pb tip and a Pb(110) surface with a dilute concentration of magnetic impurities using a dilution refrigerator STM system. By modeling the interaction of the electromagnetic environment with our Josephson STM setup, we can understand the spectroscopic data taken with a superconducting tip, including the signatures of photon-assisted inelastic Cooper pair tunneling at non-zero bias. Our measurements show a 10-15\% reduction of the Josephson critical current $I_c$ over a few Angstrom length scale in the vicinity of the magnetic adatoms, thereby demonstrating a local suppression of the order parameter. We do not observe a commensurate shift in the coherence peak energies in the single particle spectrum, which is consistent with theoretical calculations \cite{FlattePRL, FlattePRB}, highlighting the ability of the Josephson STM technique to probe physics inaccessible through traditional STM quasiparticle tunneling.  Additionally, spatially resolved spectroscopy with a superconducting tip allows us to detect novel Andreev tunneling processes through impurity-bound Shiba states of the individual adatoms.

\begin{figure}
\includegraphics[width = 9cm]{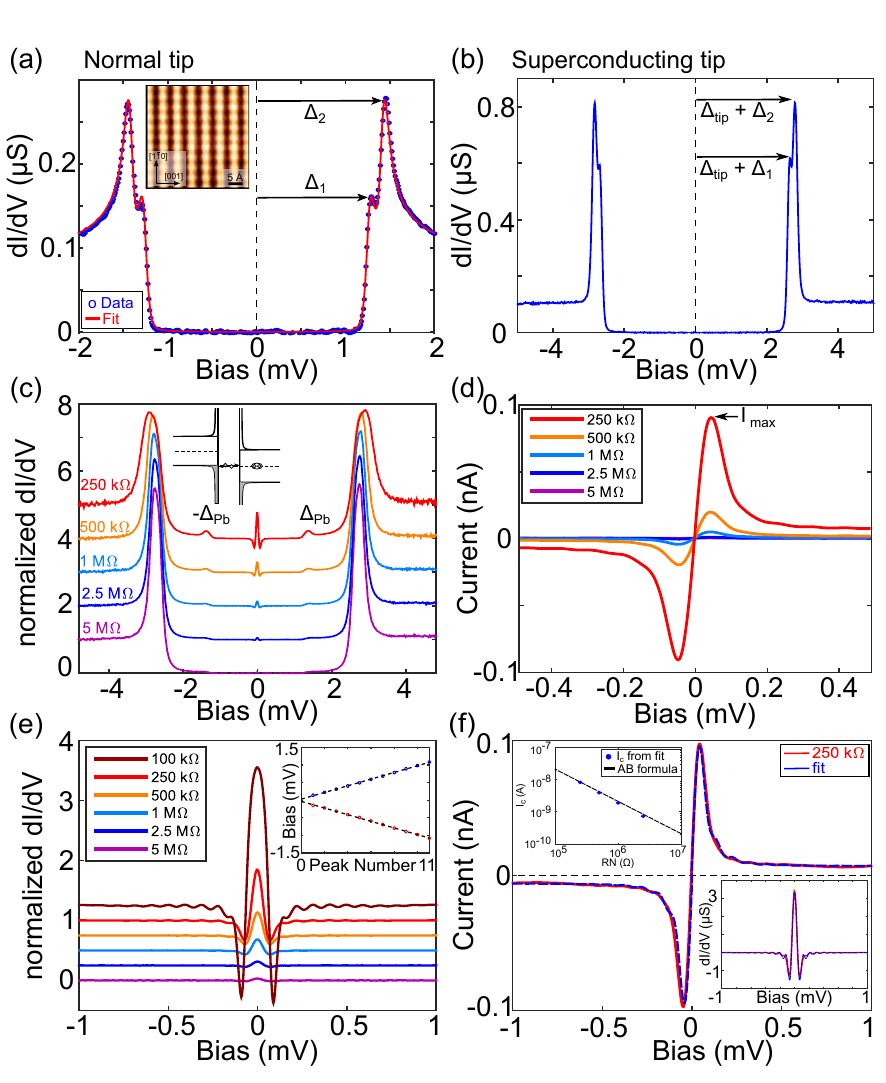}%
\caption{\label{fig1} (a) $dI/dV$ of Pb(110) with a normal tip resolving two superconducting gaps. Inset: atomic scale topography of the Pb(110) surface. (b) $dI/dV$ of Pb(110) with a superconducting Pb tip. (c) Superconducting tip spectroscopy on Pb(110), normalized to $R_N$. Curves are offset for clarity. As junction resistance is decreased, features due to Andreev reflections ($V=\pm\Delta_{Pb}$) and the Josepshon effect ($V = 0$) become more pronounced. Due to the quality of the tip, the two superconducting gaps of Pb are not clearly resolved. Inset: schematic of an Andreev process between two superconductors with the same gap, occurring at a threshold $e|V| = \Delta_{Pb}$. (d) IV characteristics of STM Josephson junction around zero bias, for different normal state resistance $R_N$. (e) Oscillations in $dI/dV$ (offset for clarity, normalized to $R_N$) due to photon-assisted Cooper pair tunneling with characteristic frequency of $\nu \approx 23$ GHz, determined by fitting the bias of the peaks as a function of oscillation number (inset). (f) Fit to IV and $dI/dV$ (bottom right inset) data for $R_N = 250$ k$\Omega$ using the P(E) theory. Top left inset: Values of critical current extracted from P(E) fits (blue dots), in agreement with the Ambegaokar-Baratoff formula (black). All point spectra in this figure were acquired at a parking bias of $V=-5$ mV at which the normal state junction resistance was determined (well outside the superconducting gap).   
}
\end{figure}

Our measurements have been carried out using a home-built dilution refrigerator STM system, with a base temperature of 20 mK and a spectroscopic resolution that corresponds to an effective electron temperature of 250 mK \cite{misra2013}.  For the present experiments, we used a Pb(110) single crystal that was prepared in-situ with several cycles of Ar sputtering and annealing to produce an atomically ordered flat surface. Fig.~\ref{fig1}(a) shows spectroscopic measurements of the atomic Pb(110) surface [inset of Fig.~\ref{fig1}(a)] measured using a normal W tip at base temperature. The two coherence peaks in the spectrum are indicative of two pairing gaps in bulk Pb associated with two different Fermi surfaces. As shown in this figure, the data can be modeled using a sum of two BCS densities of states with corresponding gaps, $\Delta_1 = 1.26$ meV and $\Delta_2 = 1.42$ meV, including an energy broadening associated with finite temperature (250 mK) and a quasi-particle lifetime (8 $\mu$eV). Although the two gaps in Pb have been previously detected in planar junctions \cite{PhysRev.128.591,PhysRev.186.397} and in STM studies using superconducting tips \cite{ruby2015}, our ability to resolve them with a normal tip demonstrates the high energy resolution afforded by the low temperature operation of our system. 

To create a Josephson junction in our STM setup, we prepare a superconducting Pb tip by indenting a W tip into the Pb(110) substrate until the spectra measured with such a tip exhibit features indicative of quasi-particle tunneling between two superconductors \cite{rodrigo2004, ruby2015}. As seen in Fig.~\ref{fig1}(b), the tunneling spectra with such tips show sharp coherence peaks at voltages corresponding to the sum of the tip and substrate superconducting gaps, with a tip gap $\Delta_{\text{tip}}$ ranging from 1.3-1.4 meV (approximately that of bulk Pb), depending on the tip. We do not resolve additional structure in the coherence peaks arising from multiple gaps in the tip, which suggests that the superconducting apex of the tip is amorphous in nature, consistent with previous  measurements \cite{ruby2015}. For simplicity, we assume below that the gap of the tip and sample are the same and define $\Delta_{\text{Pb}} \simeq (\Delta_1 + \Delta_2)/2 \simeq \Delta_{\text{tip}} \approx 1.35$ meV.

While spectroscopy at high junction resistances with the Pb tips show only signatures of quasi-particle tunneling, even at junction resistances of about 1 M$\Omega$, we begin to resolve features associated with Andreev reflections and Cooper pair tunneling between the tip and the sample [Fig.~\ref{fig1}(c)]. Successively decreasing the junction impedance highlights the evolution of distinct features at the characteristic energy of $\pm \Delta_{Pb}$, arising from Andreev processes in the STM junction \cite{heinrich2006} as illustrated in the inset of Fig.~\ref{fig1}(c). The prominent peak in conductance at zero bias corresponds to the IV characteristic shown in Fig.~\ref{fig1}(d), where the maximum Cooper pair current $I_{max}$ occurs at a voltage near zero bias ($V_p \approx 40\ \mu$V) due to a phase fluctuating Josephson supercurrent, discussed in more detail below. Moreover, we resolve periodic features in both current and conductance that appear at remarkably regular voltage intervals ($95 \ \mu$eV), which are related to the interaction of the STM Josephson junction with its electromagnetic environment [Fig.~\ref{fig1}(e)]. Finally, we also note that at low junction impedances, the large injection of quasi-particles results in an increased broadening of the coherence peaks compared to spectra at higher junction resistance [Fig.~\ref{fig1}(c)]. 

We can understand the spectroscopic features of our STM Josephson junction by comparing the magnitude of the three relevant energy scales: (i) the Josephson coupling energy between the tip and sample, $E_J = (\Delta_{Pb}/R_N)\cdot \pi \hbar /(4e^2) = \hbar/(2e)\cdot I_c$, which depends on the normal state junction resistance $R_N$ and the pairing gap $\Delta_{Pb}$, (ii) the thermal energy corresponding to our electron temperature, $k_B T=22 \ \mu$eV, and (iii) the charging energy $E_C=(2e)^2/2C$, which depends on the capacitance $C$ of the junction. For our STM junction with a typical $C \approx 4$ fF, most measurements are performed in a regime where the charging energy dominates the behavior of the Josephson junction, $E_J \lesssim k_B T \ll E_C$. In this case, the thermal fluctuation of the phase across the junction are enhanced by quantum fluctuations, which together result in a shift of the Josephson pair current to non-zero bias \cite{PhysRevB.50.395}, as we observe in our data shown in Fig.~\ref{fig1}(d). The oscillations in the conductance at higher bias [Fig.~\ref{fig1}(e)] arise from the interaction between such a phase incoherent Josephson junction and the standing electromagnetic modes associated with the STM tip. Photon absorption or emission with energy $E = h\nu = 2eV$ facilitate Cooper pair tunneling across the STM junction at non-zero bias, a signature similar to the AC Josephson effect \cite{jack2015,roychowdhury2015}. Treating our STM setup as an open-ended $\lambda/4$ antenna with resonances at $\nu_n = (2n+1)\cdot c/l$ \cite{jack2015} allows us to fit the voltage spacing of the oscillations to extract a characteristic frequency $\nu_0 \approx 23$ GHz, which is intrinsic to our instrument, corresponding to an approximate tip length $l=3.3$ mm. The high quality factor of these oscillations highlight the sensitivity of the STM Josephson junction to its electromagnetic environment.

\begin{figure*}
\includegraphics[width = \textwidth]{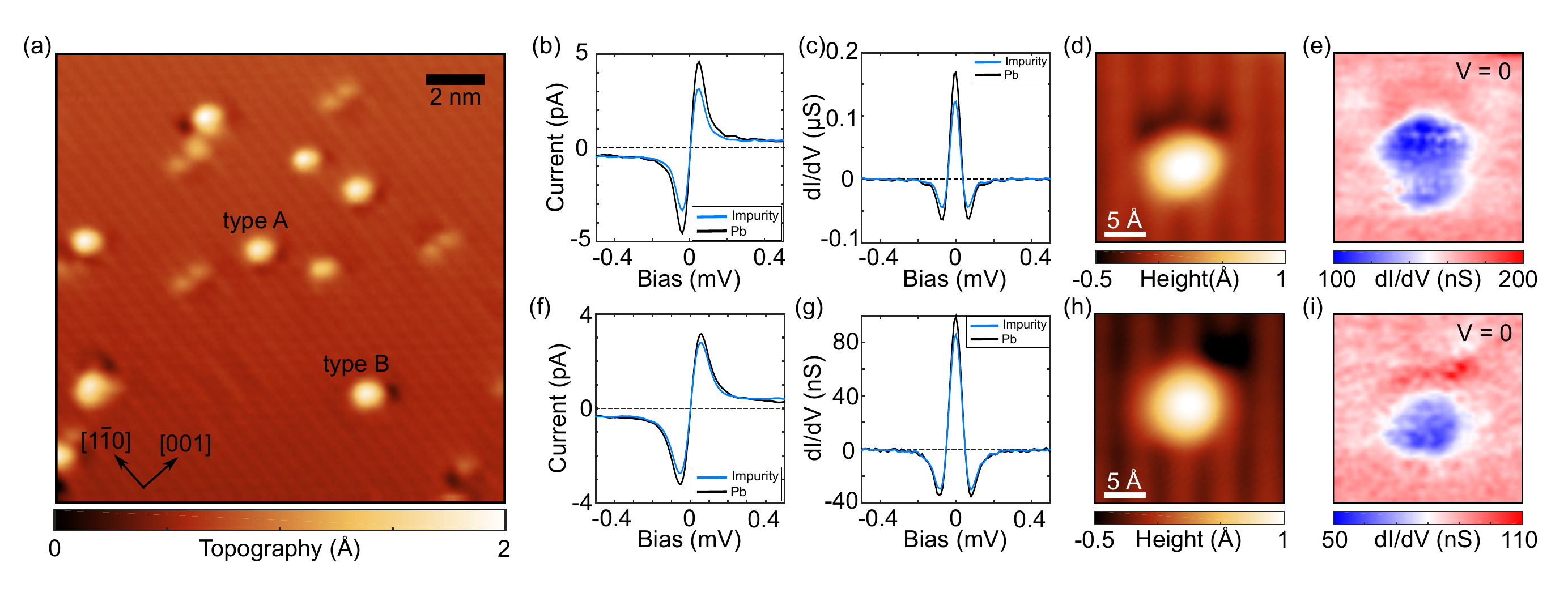}%
\caption{\label{fig2} (a) Topography of Pb after evaporating a submonolayer of Fe. Individual Fe adatoms tend to be centered on a trench (type A) or row (type B) of the Pb(110) surface as marked. Suppression of pair current [(b),(f)] and $dI/dV$ [(c),(g)] on the Fe adatom (blue) compared to on Pb (black) at 1 M$\Omega$ junction impedance for type A [(b)-(e)] and type B [(f)-(i)] adatoms. Simultaneous topographies [(d),(h)] and differential conductance maps [(e),(i)] at $V=0$ for $R_N = 1$ M$\Omega$ demonstrate the spatial suppression of the Josephson effect over the Fe adataoms. Parking conditions $V = -5$ mV and $I = 5$ nA for data in (b)-(i), ensuring that $R_N$ is the same over both the adatom and bare Pb. Different superconducting tips were used for the point spectra and the conductance maps, but the $I_c$ suppression is consistent across measurements.}
\end{figure*}

A quantitative understanding of our Josephson STM characteristics can be obtained by using the P(E) theory to model the probability of Cooper pair tunneling across the junction mediated by its electromagnetic environment \cite{ingoldGrabert1991,PhysRevB.50.395}. The energy exchange with the environment as well as thermal effects contribute to inelastic pair tunneling across the junction. As shown in the fit in Fig.~\ref{fig1}(f), we can accurately capture the particular shape of our spectra using this theory, the details of which are discussed in the Supplementary Materials. Within this model, both the maximal phase-fluctuating pair current $I_{max}$ and the differential conductance at zero bias $dI/dV(V=0)$ are proportional to the square of the intrinsic Josephson critical current, $I_c^2$ \cite{ingoldGrabert1991,PhysRevB.50.395}. The values that we obtain for the intrinsic $I_c$ as a function of the normal state junction resistance $R_N$ are in good agreement with calculations from the Ambegaokar-Baratoff formula, $I_c = \pi/(2e)\cdot \Delta_{Pb}/R_N$ \cite{ambegaokar1963}, as shown in the top inset of Fig.~\ref{fig1}(f), justifying the use of this model to fit the data. Although the small capacitance of a typical STM junction makes such Josephson junctions phase incoherent, mapping of the phase-fluctuating pair current at low bias in the STM setup still provides a direct method for the spatial characterization of the pairing amplitude in a superconducting sample. 

To demonstrate that the Josephson STM technique can probe variations of the order parameter on the atomic scale, we investigate individual magnetic atoms deposited on the surface of our Pb(110) substrate. Magnetic impurities on an s-wave superconductor are the simplest example of pair-breaking defects that are well known to suppress superconductivity, for instance seen in the suppression of $T_c$ with increasing impurity concentration \cite{anderson1959,abrikosov1961}. Previous STM spectroscopy has used quasiparticle tunneling to show that magnetic atoms on a BCS superconductor induce in-gap Shiba states \cite{shiba1968,yu1965,rusinov1969,yazdani1997,PhysRevLett.100.226801}; however such LDOS measurements do not directly probe the superconducting order parameter. To probe the spatial variation of the pairing strength for this model system, we deposit a sub-monolayer of Fe adatoms on Pb(110) in situ, at a temperature of about 20 K. As shown in a typical STM topography [Fig.~\ref{fig2}(a)], this low temperature deposition results in the appearance of features with Angstrom height, which are consistent with individual Fe atoms residing in two different atomic sub-lattice binding sites (types A and B) on the Pb(110) substrate [Figs.~\ref{fig2}(d),(h)].

To probe the pairing amplitude near these magnetic Fe atoms, we perform spectroscopic measurements using superconducting Pb tips at junction resistances that are low enough to detect the phase fluctuating Josephson current but high enough to perform STM imaging without disrupting the adatoms on the surface. As shown in Figs.~\ref{fig2}(b),(f), IV measurements of the two different types of Fe adatoms sites show a suppression of the peak in the pair current $I_{max}$ near zero bias as compared to the bare substrate, also seen in an analogous measurement of differential conductance over the magnetic impurities [Figs.~\ref{fig2}(c),(g)]. This reduction of the phase-fluctuating pair current and conductance is a direct signature of the local suppression of the pairing amplitude caused by the magnetic impurities, corresponding to a 10-15\% reduction of the critical current $I_c$ on an Fe adatom based on fits to the spectra (see Supplementary Materials). 
The precise value of $I_{max}$ and the zero bias conductance depend on the superconducting tips and the strength of the exchange coupling between the impurity and underlying Pb substrate (as is evident from the difference between type A and B adatoms), but the suppression of the intrinsic critical current is consistent across dozens of adatoms measured with various different Pb tips.  

\begin{figure}
\includegraphics[width = \columnwidth]{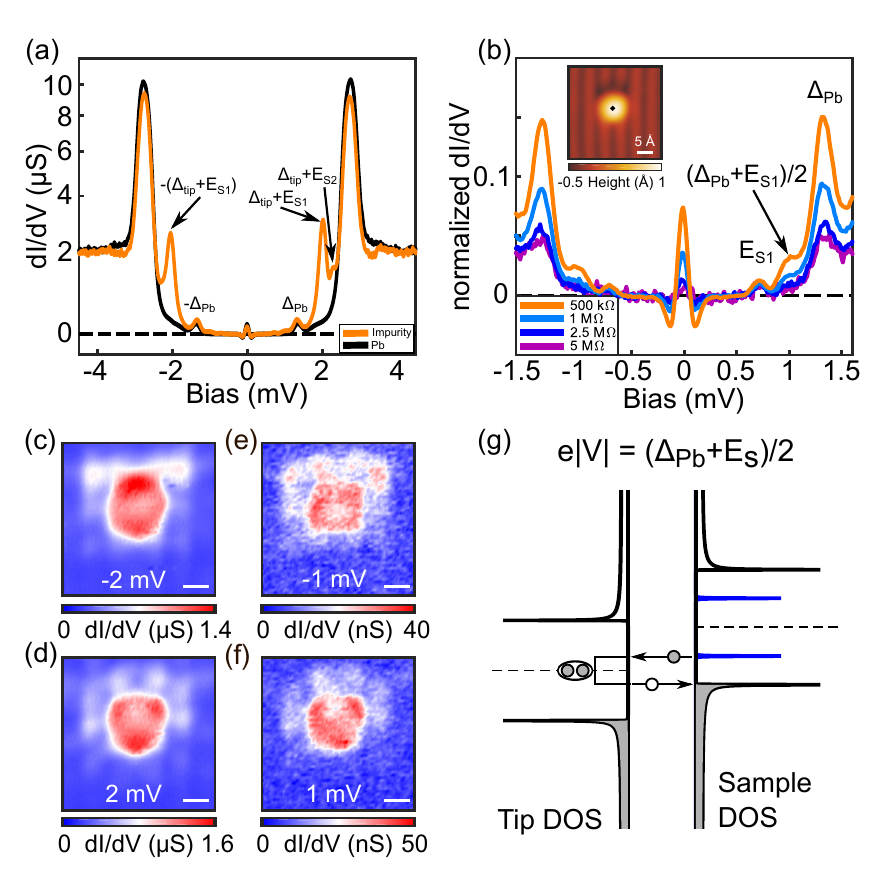}%
\caption{\label{fig3} (a) $dI/dV$ with a superconducting tip ($V=-5$ mV, $I=10$ nA) on a type A impurity, plotted on a log scale, showing the appearance of Shiba states at $\Delta_{Pb}+E_{S1} = \pm2.06$ meV and $\Delta_{Pb}+E_{S2}=+2.28$ meV (marked by arrows). (b) A zoom-in of $dI/dV$ at lower junction resistances highlights multiple Andreev reflections through the Shiba state observed at energies of $E_{S1}$ and $(\Delta_{Pb}+E_{S1})/2$. Parking bias $V=-5$ mV for all spectra. [(c)-(f)] Conductance maps at $eV=\pm(\Delta_{Pb}+E_{S1})$ and $eV=\pm(\Delta_{Pb}+E_{S1})/2$ taken simultaneously with topograph in inset of (b). Corresponding spatial patterns between the Shiba state ($V=2$ mV) and the Andreev reflection state ($V=1$ mV), and similarly for states at negative bias are observed. Scale bar is 5 $\mathrm{\AA}$. (g) Schematic for Andreev reflection process with threshold voltage $eV=\pm(\Delta_{Pb}+E_{S1})/2$, where the relevant energy gap is that of the sample.}
\end{figure}

Another measurement that reflects the local pairing suppression is mapping the differential conductance at zero bias in the vicinity of the Fe defects to probe the spatial variations of the order parameter. These measurements, shown in Figs.~\ref{fig2}(e),(i), demonstrate our key finding that this Josephson STM technique can resolve spatial variations of the pairing amplitude on the atomic scale. Data from additional adatoms showing a suppression of the order parameter, as well as control experiments that rule out artifacts of tip height variations are detailed in the Supplementary Materials.
Our observation that the pairing amplitude recovers back to its unperturbed value for the Pb within Angstroms of the magnetic adatoms, and not on the scale of the coherence length is consistent with theoretical predictions \cite{heinrichs1968,schlottmann1976}. Theory also predicts an oscillating power-law for the order parameter, $\Delta (r) \sim \sin^2(k_Fr)/(k_Fr)^2$ as a function of the distance $r$ from a magnetic impurity, where $k_F$ is the Fermi wavevector \cite{schlottmann1976}. However, both the adatom geometry on the surface of a 3D superconductor and the short Fermi wavelength in Pb give rise to the short-range decay of the order parameter, making it difficult to detect these oscillations in the current experiment.

In contrast, the suppression of the local order parameter is not prominently seen in quasiparticle tunneling measurements over the magnetic impurity. Superconducting tip spectra [Fig.~\ref{fig3}(a)] over an Fe adatom show a suppression in the intensity of the coherence peaks with no appreciable shift in their energy as compared to the bare substrate. In the presence of bound states, the spectral weight is redistributed from the coherence peaks to the bound state energies, thus the suppression of the local pairing amplitude does not translate into a shift in the energies of the coherence peaks. Self-consistent calculations confirm this distinction between measurements of the order parameter and the local density of states \cite{FlattePRL, FlattePRB}. 
Thus, our measurements demonstrate the utility of scanning Josephson spectroscopy in directly extracting local variations of the superconducting order parameter.

Spectroscopic measurements with a superconducting tip can also be used to probe previously unexplored Andreev reflection processes through the in-gap Shiba states localized near individual magnetic adatoms. Typically, impurity-bound Shiba states are detected in measurements with superconducting tips through quasi-particle or Andreev tunneling processes at an energy of $eV=\Delta_{\text{tip}}+E_S$, where $E_S$ is the energy of the Shiba state \cite{PhysRevLett.100.226801,PhysRevLett.115.087001}. Performing similar measurements on our Fe defects, we find the expected signatures of the Shiba states, an example of which is shown in Fig.~\ref{fig3}(a) for a type A Fe defect, resolving two different Shiba states at $e|V|=\Delta_{\text{tip}}+E_{S1}$ and $e|V| = \Delta_{\text{tip}}+E_{S2}$. Examining the dependence of such measurements on the coupling between the superconducting tip and sample, we find that at lower junction resistance, there are additional subgap features in the spectra at lower biases [Fig.~\ref{fig3}(b)]. In particular, we find peaks in conductance at $e|V|=E_{S1}$ and $e|V| = (\Delta_{Pb}+E_{S1})/2$, which correspond to previously undetected Andreev reflections through an impurity-bound Shiba state. The feature at $e|V| = (\Delta_{Pb}+E_{S1})/2$, which is due to a sub-gap tunneling process, cannot occur due to quasiparticle tunneling but rather arises from an Andreev process through the Shiba state, as illustrated in Fig.~\ref{fig3}(g). Further corroboration that such Andreev processes involve the localized Shiba state can be obtained by comparing the spatial patterns of maps over the impurities at the Andreev reflection energy ($e|V| = (\Delta_{Pb}+E_{S1})/2$) with that of tunneling at $e|V| = \Delta_{Pb}+E_{S1}$. The excellent correspondence between the hole-like (electron-like) conductance measurements shown in Figs.~\ref{fig3}(c),(e) (Figs.~\ref{fig3}(d),(f)) for the type A impurity demonstrates the role of Shiba states in such Andreev processes. (See also Supplementary Material for other examples and a discussion of possible complications due to imperfect superconducting tips). 

In conclusion, we demonstrate that combining Josephson spectroscopy with atomic resolution STM imaging provides a method to directly probe the local superconducting order parameter, which is not directly accessible through traditional quasiparticle measurements. Despite the challenge of operating such Josephson junctions in a fully phase coherent regime due to their ultra small dimensions, the fluctuating pair current is still a powerful tool to examine spatial variations of the pairing amplitude. Our key accomplishment, demonstrating that these measurements can be performed with atomic resolution, paves the way for using scanning Josephson techniques to study an inhomogeneous or spatially modulated order parameter in novel superconducting materials.

\section{Acknowledgements}
\begin{acknowledgments}
We would like to thank J. Li for valuable discussions. This work has been supported by the Gordon and Betty Moore Foundation as part of EPiQS initiative (GBMF4530) and DOE-BES. This project was also made possible using the facilities at Princeton Nanoscale Microscopy Laboratory supported by grants through NSF-DMR-1104612, ARO-W911NF-1-0262, ONR-N00014-14-1-0330, ONR-N00014-13-10661, DARPA-SPWAR Meso program N6601-11-1-4110, LPS and ARO-W911NF-1-0606, and NSF-MRSEC programs through the Princeton Center for Complex Materials DMR-1420541. MTR acknowledges support from the NSF Graduate Research Fellowship. 
\end{acknowledgments}

\bibliography{josephson_resub}{josephson_resub}{}

\pagebreak
\begin{center}
\textbf{\large Supplementary Materials: Scanning Josephson spectroscopy on the atomic scale}
\end{center}

\setcounter{equation}{0}
\setcounter{figure}{0}
\setcounter{table}{0}
\setcounter{page}{1}
\renewcommand{\thefigure}{S\arabic{figure}}
\renewcommand{\figurename}{FIG.}

\section{Fitting procedure for Josephson critical current}

The general tunneling properties of ultrasmall junctions are well described by the P(E) theory, which can be tailored to the case of Cooper pair tunneling between two superconducting junctions by treating the Josephson energy as a perturbation to the charging energy ($E_J \ll E_C$) \cite{supp_PhysRevLett.64.1824,supp_ingoldGrabert1991,supp_PhysRevB.50.395}. This theory considers a voltage-biased junction coupled to an arbitrary environment, where inelastic tunneling processes mediated by the external environment are characterized by a probability distribution $P(E)$. For STM junctions with a capacitance on the order of 1 fF, the dynamics are dominated by the charging energy, which gives rise to quantum fluctuations of the phase. However, photon assisted tunneling of Cooper pairs is still possible if the environment is able to compensate for the bias by emitting or absorbing a photon of energy $E=2eV=h\nu$ as a Cooper pair tunnels across the junction. 
Except in the case of simple environmental impedance functions, it is difficult to obtain an analytic expression for P(E). We follow a numerical method to compute the probability of Cooper pair tunneling due to interactions with an environment modeled by a complex impedance $Z(\omega)$, as discussed in \cite{supp_ingoldGrabert1991}, using the convolution
\begin{equation}
P_Z(E) = I(E) + \int^{+\infty}_{-\infty} d\omega K(E,\omega) P_Z(E-\hbar\omega). 
\end{equation}
The inhomogeneity $I(E)$ is given by a Lorentzian representing the effect of the Ohmic resistance of the environment and serves as an initial guess for P(E). Detailed forms of $I(E)$ and $K(E,\omega)$ are given in \cite{supp_ingoldGrabert1991}. We solve this convolution iteratively until self consistency is reached. 

\begin{figure}
  \centering
  \includegraphics[width=15cm]{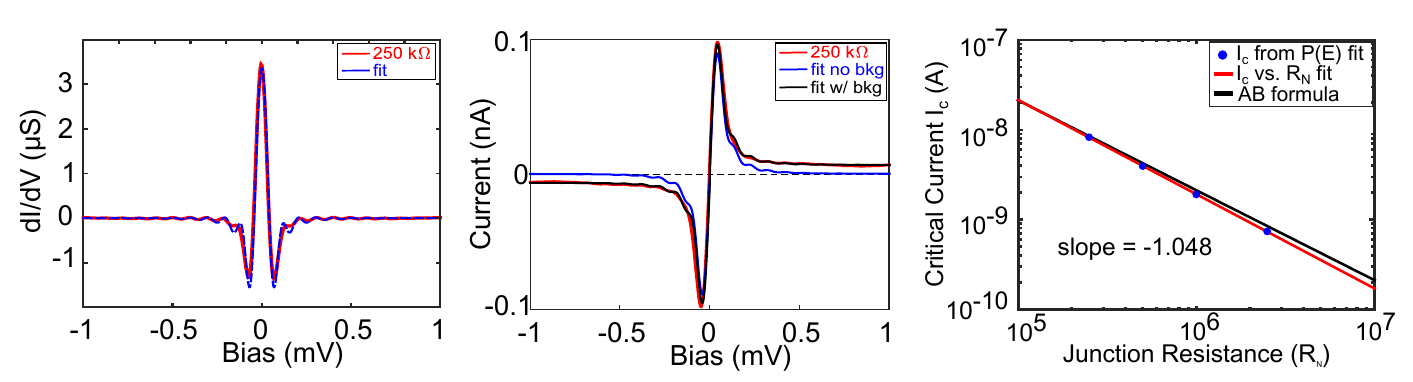}
  \caption{Here we provide additional details of the fit shown in Fig. 1(f) of the main text. (a) Fit to $dI/dV$ with a Pb tip on a Pb(110) sample for $R_N = 250$ k$\Omega$ based on the numerical implementation of the P(E) theory. Fitting parameters are $C=4.3$ fF, $R_L = 502 \ \Omega$ and $\alpha = 0.15$ with the temperature fixed at $T=250$ mK. From the overall amplitude of the fit, we obtain a critical current $I_c = 8.25$ nA compared to $I_c = 8.5$ nA from the Ambegaokar-Baratoff formula for $\Delta_{Pb} = 1.35$ meV. (b) Calculated I(V) from the $dI/dV$ fitting parameters (blue) demonstrates the need to include an additional term to account for a quasiparticle background in order to match the 250 k$\Omega$ I(V) data. A constant term, which of the form of the form $c \cdot 2(\Theta(V)-0.5)$ where $\Theta(x)$ is the Heaviside function, is added with $c=6.5$ pA (black). (c) For critical currents extracted from the fitting procedure, we find $I_c \propto R_N^{-1.048}$, with the values of $I_c$ in good agreement with the Ambegaokar Baratoff formula. }
  \label{fig:PEfitting}
\end{figure}

The data show periodic oscillations in the differential conductance [Fig.~\ref{fig:PEfitting}(a)], which we presume arise from standing modes of the tip. We model the STM tip geometry as an open-ended finite transmission line, formed between the tip holder and the sample, with eigenfrequencies corresponding to $\nu_n = (2n+1)\cdot c/l$, for an STM tip of length $l$, where $c$ is the speed of light \cite{supp_jack2015}. 
The total impedance seen by the junction is given according to \cite{supp_PhysRevB.50.395}
\begin{equation}
\frac{Z_t(\omega)}{R_Q} = \rho \frac{1+\frac{i}{\alpha} \tan \left( \frac{\pi}{2} \frac{\omega}{\omega_0} \right)}{\left[1-\delta \frac{\omega}{\omega_0}\tan \left( \frac{\pi}{2} \frac{\omega}{\omega_0} \right) \right]+ i\alpha \left[ \delta \frac{\omega}{\omega_0}+ \tan \left( \frac{\pi}{2} \frac{\omega}{\omega_0} \right) \right]},
\end{equation}
where $\rho = R_L/R_Q$ for a load resistance $R_L$ and the quantum of resistance $R_Q = h/(2e)^2 = 6.45 k\Omega$, $\omega_0 = 2\pi\nu_0$ is the characteristic $\lambda/4$ frequency, $1/\alpha$ is an effective quality factor of the resonator and $\delta = \omega_0RC$, the ratio of the $\lambda/4$ frequency to the cutoff frequency $1/RC$. The load resistance provides a measure of the effective ohmic impedance seen by the junction, which is on the order of the vacuum impedance $Z_0 = 377 \Omega$, largely determined by the wiring.  

The dissipative interaction of a quantum system with the environment opens up one path for photon assisted tunneling as discussed above. Thermal fluctuations on the capacitive junction create an additional channel for inelastic tunneling of Cooper pairs, where the voltage noise across the capacitor has an rms $\sqrt{\bar{v}^2} = \sqrt{k_BT/C}$, corresponding to a gaussian distribution \cite{supp_jack2015}
\begin{equation}
P_C(E) = \frac{1}{\sqrt{4\pi E_C k_B T}}\exp \left[ - \frac{E^2}{4 E_C k_B T}  \right].
\end{equation}

The total $P(E)$ is the convolution of the probability distributions from these two channels $P_Z(E)$ and $P_C(E)$.
The net Cooper pair current is given by 
\begin{equation}
I(V) = \frac{\pi e E_J^2}{\hbar}[P(2eV)-P(-2eV)] = \frac{\pi \hbar I_c^2}{4e}[P(2eV)-P(-2eV)],
\label{eq1}
\end{equation}
where the Josephson tunneling determines the overall amplitude which is proportional to $I_c^2$ \cite{supp_ingoldGrabert1991,supp_PhysRevB.50.395}.
The fitting parameters we include in our numerical model are (i) a junction capacitance $C$, (ii) a load resistance $R_L$, (iii) a characteristic frequency $\nu_0$, which is intrinsic to the system and can be used to determined the length of the STM tip (iv) $\alpha$, which is inversely proportional to the quality factor of the resonator and (v) an overall amplitude $A = (\pi \hbar)/(4e)\cdot I_c^2$ from which the critical current can be extracted. We fix the temperature to the electron temperature $T = 250$ mK, which accounts for extrinsic sources of noise (e.g. from room temperature electronics; for details of the measurement circuit, see \cite{supp_misra2013}). Since the resonances of the tip eigenmodes are more clearly resolved in the differential conductance for tunneling between a Pb tip and a Pb(110) sample, we fit $dI/dV$ to the derivative of current calculated from the P(E) theory [Fig.~\ref{fig:PEfitting}(a)]. In order to fit the IV characteristics of the junction, an additional constant term must be added to account for the background [Fig.~\ref{fig:PEfitting}(b)], likely arising from quasiparticle poisoning. 

Fig.~\ref{fig:PEfitting}(c) shows the linear dependence of the critical current values extracted from the fit of $dI/dV$ over a range of junction resistances $R_N$. This is in remarkable agreement with the Ambegaokar Baratoff formula \cite{supp_ambegaokar1963}
\begin{equation}
I_c = \frac{\pi}{2e} \frac{\Delta_{Pb}}{R_N},
\end{equation}
 assuming the gap is the same for the tip and the sample with $\Delta_{Pb} = 1.35$ meV. Thus, even though we are in the phase fluctuating regime, the incoherent Cooper pair tunneling across our Josephson STM junction provides a measure of the local order parameter of the system. The maps in Figs. 2 (main text) and ~\ref{fig:setpointEffect} show the zero bias conductance on Pb to be constant to within in 2-3\% (1 standard deviation) of the average value. 
For measurements performed with the same superconducting tip, the values of the remaining fit parameters, $C, R_L, \nu_0, \alpha$, have a less than 5\% variation at different junction resistances between 2.5 M$\Omega$ and 250 k$\Omega$. For different superconducting tips, the capacitance and resistance show a 10-15\% variation, whereas $\alpha$ can vary by up to 30\%. 

\section{Suppression of the Josephson effect over Fe adatoms}
\begin{figure}
  \centering
  \includegraphics[width=13cm]{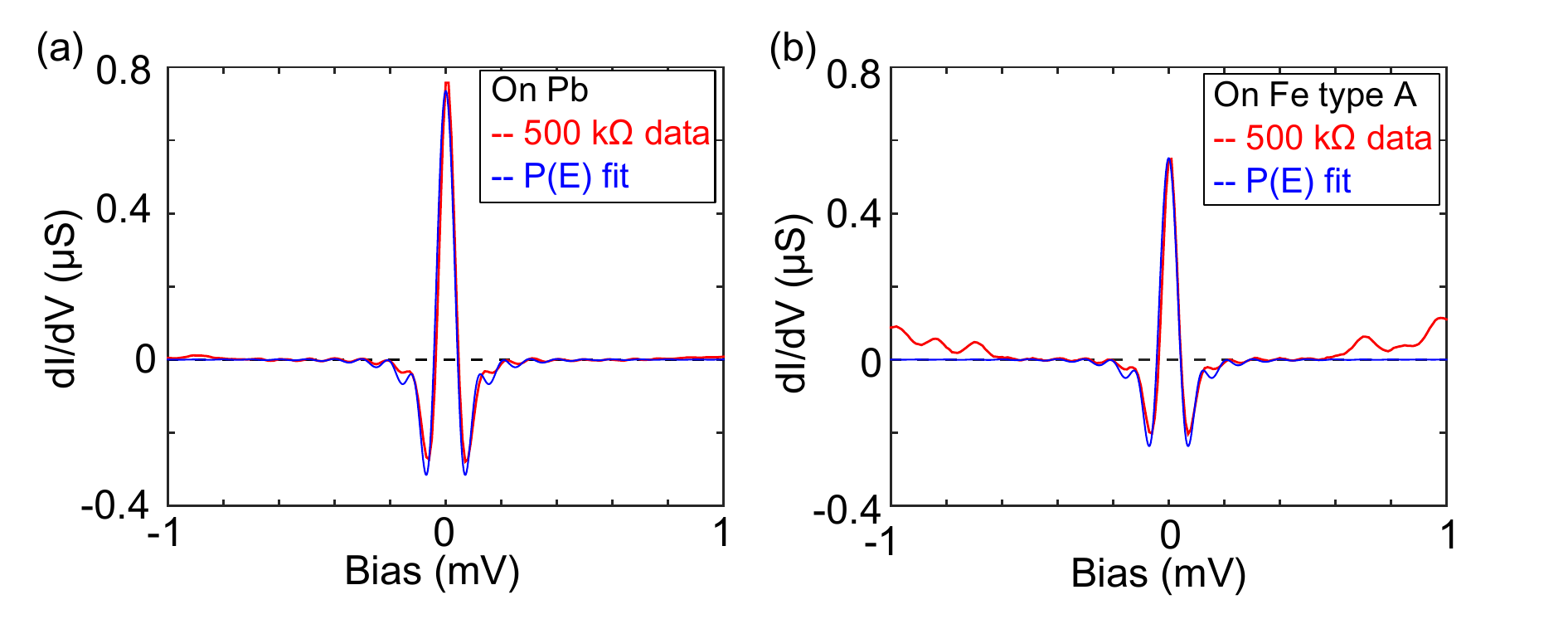}
  \caption{Fits to differential conductance on (110) surface of Pb compared to that on an Fe adatom showing a suppression of the Josephson effect. Data measured at a junction resistance $R_N = 500$ k$\Omega$ with the same superconducting Pb tip. Fit parameters for both panels: $C=4.1$ fF, $R_L = 550 \ \Omega$ and $\alpha = 0.18$. From the overall fit parameter, we obtain values for the critical current, $I_c = 4.15$ nA on Pb compared to $I_c = 3.60$ nA on the Fe adatom, indicating a 13\% suppression over the magnetic impurity.}
  \label{fig:IcSuppression}
\end{figure}

A quantitative treatment of the suppression of the Josephson critical current over an Fe adatom can be obtained by comparing fits to the conductance over the impurity to that on clean Pb. Fig.~\ref{fig:IcSuppression}(a)-(b) shows an example fit to $dI/dV$ for Pb and a type A impurity measured with the same superconducting tip, demonstrating a 13\% suppression of $I_c$. Overall, we find a 10-15\%  suppression of the critical current over the Fe impurity for dozens of adatoms (both type A and type B) measured with various different superconducting tips at several junction resistances.  

Next, we present data from two additional adatoms and discuss several measurements that rule out variations in tip-sample distance as a cause for the suppression of the critical current. We simultaneously establish the junction resistance well outside the gap and perform spectroscopic measurements of the Josephson effect on the Fe adatoms. The parking conditions (setpoint current and voltage) that determine $R_N$ are identical both on the bare Pb as well as over the Fe adatoms, ensuring there is no positional variation in $R_N$. Moreover, we note that the magnitude of suppression we observe makes it highly unlikely that the change in z-height is responsible for the reduction of the Josephson effect. Since the tunneling current is exponentially sensitive to tip-sample distance, a 1 $\mathrm{\AA}$ increase in tip height would lead to an order of magnitude decrease in the intrinsic Josephson critical current. In contrast, we observe only a 10-15\% suppression of the Josephson critical current over the Fe adatoms, which shows that it is not related to an increased distance from the Pb(110) surface.

\begin{figure}
  \centering
  \includegraphics[width=15cm]{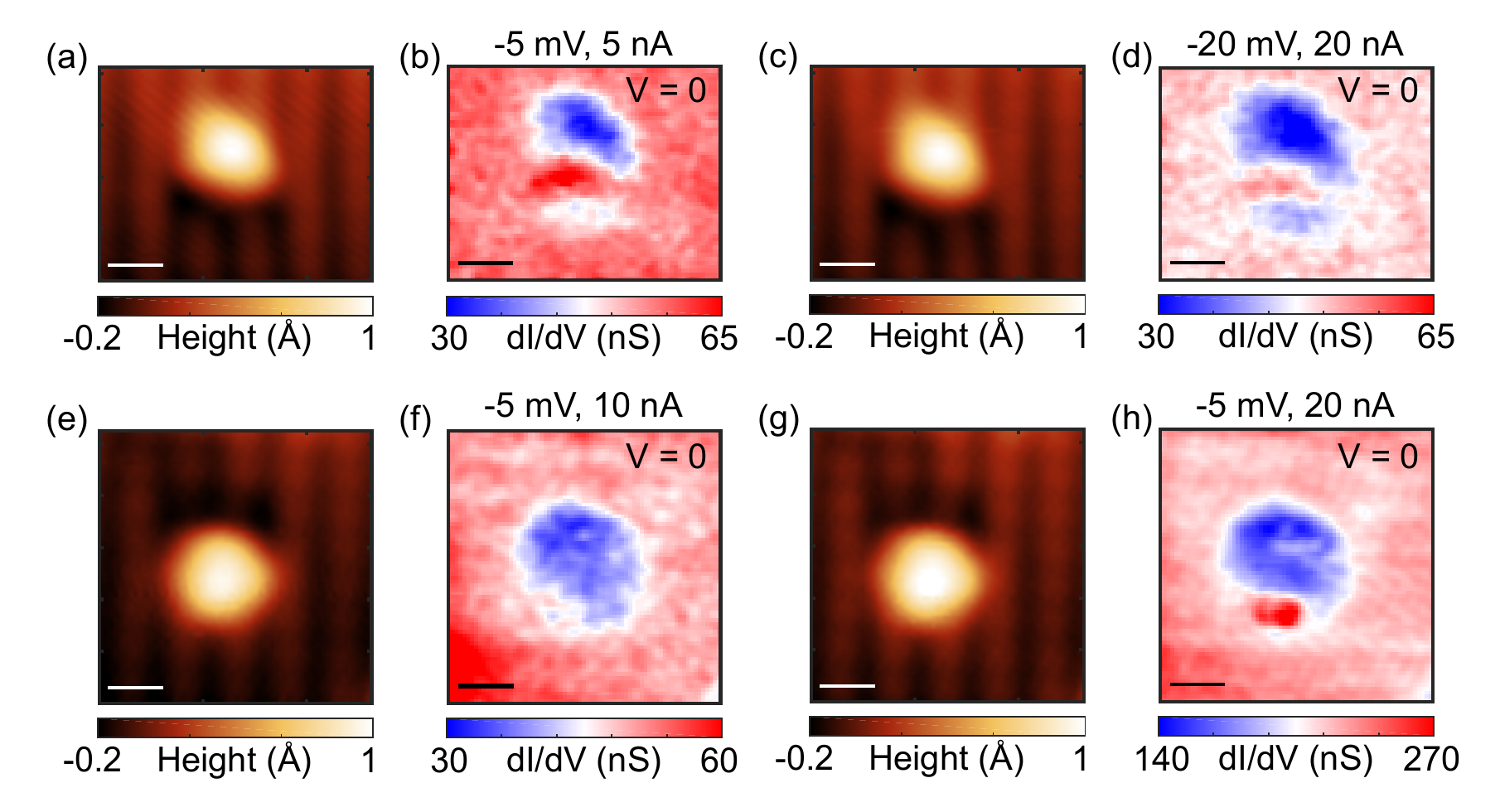}
  \caption{(a)-(d) Simultaneous topographies and differential conductance maps at $V=0$ for $R_N = 1 \ \mathrm{M\Omega}$ over an Fe adatom with the same tip for different parking conditions: $V = -5$ mV, $I = 5$ nA [(a)-(b)] and $V = -20$ mV, $I = 20$ nA [(c)-(d)]. In both cases, the spatial pattern of the suppression of the zero bias conductance is identical, and the slight discrepancies in absolute conductance values is likely due to the phonon modes in Pb. (e)-(h) Simultaneous topographies and differential conductance maps over a second Fe adatom with the same tip (but different than that used for (a)-(d)) for different values of junction resistance, showing similar spatial patterns of suppression. (e)-(f) Parking conditions of $V=-5$ mV, $I=10$ nA corresponding to $R_N = 500$ k$\mathrm{\Omega}$. (g)-(h) Parking conditions of $V=-5$ mV, $I=20$ nA corresponding to $R_N = 250$ k$\mathrm{\Omega}$. Scale bar in all panels is 5 $\mathrm{\AA}$. }
  \label{fig:setpointEffect}
\end{figure}

Spectroscopic maps of the zero bias differential conductance at $R_N = 1 \ \mathrm{M\Omega}$, taken with the same superconducting tip at two different parking biases of $V=-5 \ \mathrm{mV}$ and $V=-20 \ \mathrm{mV}$ are shown in Fig.~\ref{fig:setpointEffect}(a)-(d). In order to maintain the same junction resistance, the setpoint current is four times as large in the second measurement ($I = 5$ nA and $I = 20$ nA, respectively). Both data sets show similar conductance values and identical spatial patterns for zero bias suppression, confirming that the $I_c$ suppression is not due to spurious signals arising from differences in the density of states outside the gap over the bare Pb compared to the adatoms. The slight discrepancy in absolute conductance values is likely due to the fact that the $V=-20$ mV parking bias is well outside strong-coupling energy regime in Pb, whereas the phonon modes are still present at $V=-5$ mV. Fig.~\ref{fig:setpointEffect}(e)-(h) compares zero bias differential conductance maps taken at the same parking bias of V = -5 mV for two different junction resistances, $R_N$ = 500 k$\mathrm{\Omega}$ and $R_N$ = 250 k$\mathrm{\Omega}$, corresponding to a difference in tip-sample distance between the two data sets, and a decrease in tip height relative to the 1 M$\mathrm{\Omega}$ measurements. The two measurements show similar spatial patterns and a similar percentage suppression of the zero bias conductance over the adatom, though the absolute conductance values differ by the expected amount based on $R_N$. Given the exponential sensitivity of tunneling current to tip height, a factor of four variation in the junction impedance is considerable. Therefore, we rule out measurement artifacts from z-height variations and attribute the reduction in $I_c$ to the suppression of the superconducting order parameter over the magnetic impurities.

\section{Andreev reflections through impurity-bound Shiba states}

\begin{figure}
  \centering
  \includegraphics[width=10cm]{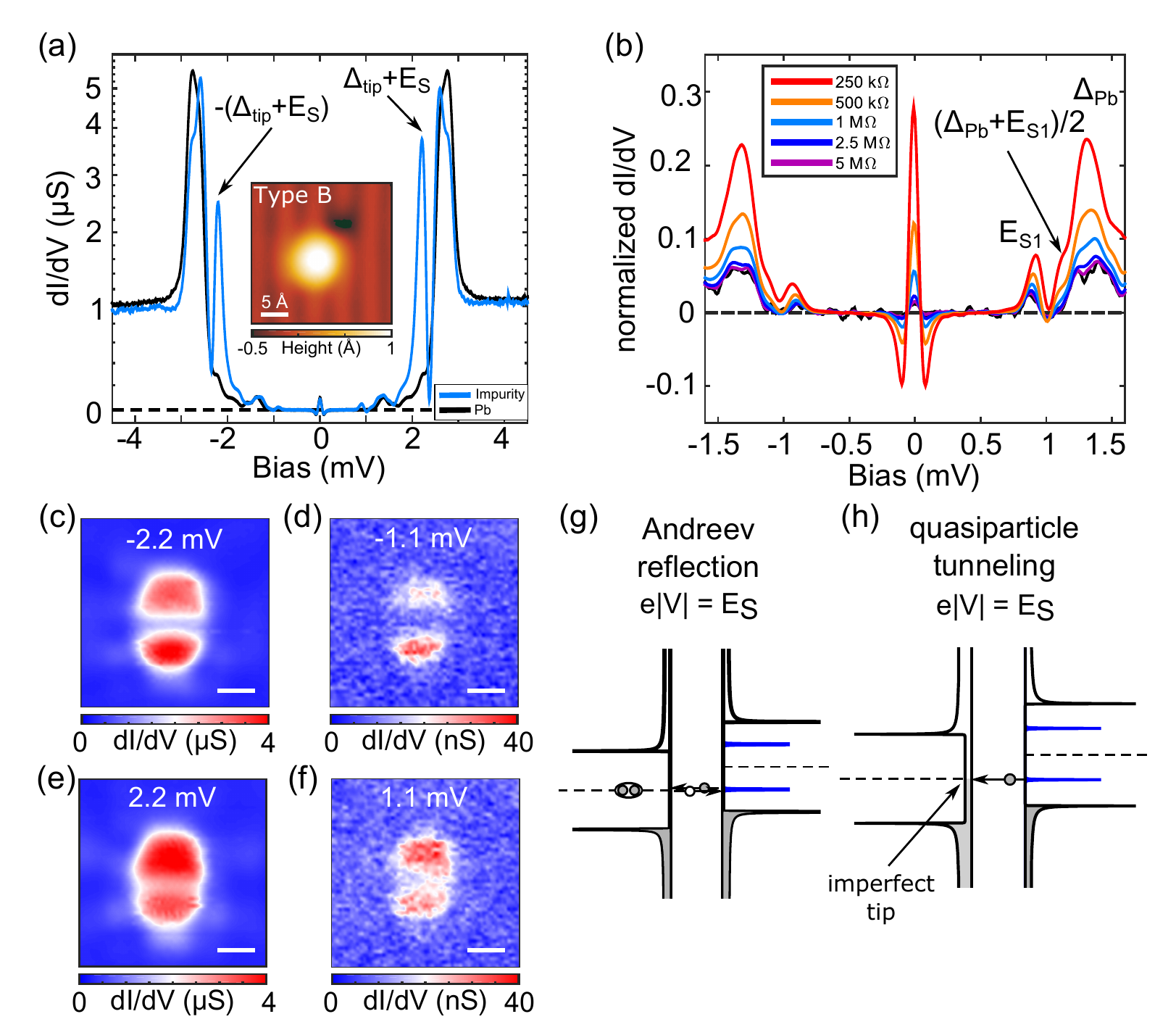}
  \caption{(a) Superconducting tip spectra on a type B Fe impurity demonstrating the presence of Shiba states. Measured at $R_N = 1\ M\Omega$ ($V=-5$ mV, $I=5$ nA), plotted on a log scale. (b) Point spectra on a type B Fe adatom showing sub-$\Delta$ features at $E_S$ and $(\Delta_{Pb}+E_S)/2$, which increase in strength as junction resistance is decreased. Parking bias $V=-5$ mV, normalized to $R_N$. [(c)-(f)] Conductance maps at $eV = \pm (\Delta_{Pb}+E_S)$ and $eV = \pm (\Delta_{Pb}+E_S)/2$ with corresponding spatial patterns between the Shiba state ($V$ = 2.2 mV) and the sub gap Andreev reflection feature ($V$ = 1.1mV), and similarly for negative bias. Scale bar is 5 $\mathrm{\AA}$. (g) Andreev reflection process which would give rise to a peak in $dI/dV$ at the bare Shiba state energy $E_S$ that is expected to get stronger for lower junction resistance. (h) Direct quasiparticle tunneling into the Shiba state due to an imperfect tip could also give rise to a peak in $dI/dV$ at $E_S$, which would be independent of junction resistance.}
  \label{fig:typeBshiba}
\end{figure}

Here we provide additional data on Andreev reflections through the localized Shiba states of Fe adatoms on a Pb(110) surface. The Shiba states measured with a superconducting tip for the type B Fe adatom are shown in Fig.~\ref{fig:typeBshiba}(a), where the presence of the impurity also changes the coupling of the tip into the two gaps of Pb(110). Fig.~\ref{fig:typeBshiba}(b) shows the evolution of subgap Andreev processes as the junction resistance is decreased. The Andreev reflections that give rise to the peaks at $eV = \pm \Delta_{Pb}$ and $eV = \pm (\Delta_{Pb}+E_S)/2$ are discussed in the main text [inset of Fig. 1(c) and Fig. 3(g)]. 

The schematic in Fig.~\ref{fig:typeBshiba}(g)-(h) depicts two processes which could give rise to the feature in the conductance at $e|V| = E_S$. An Andreev process at the Shiba state energy would involve an electron (hole) sourced from the Shiba state to be retroreflected as a hole (electron) back into the same state. On the other hand, an imperfect tip with some non-zero DOS up to $E_F$ could give rise to direct quasiparticle tunneling into the Shiba state. We expect that an Andreev process should get stronger for larger tip-sample coupling (smaller junction resistance), but quasiparticle tunneling is expected to be independent of $R_N$. For the type A adatom [main text Fig. 3(b)], the peak in $dI/dV$ at $E_S$ persists even for high resistances ($R_N >1M\Omega$), an indication that the peak arises from a combination of quasiparticle and Andreev tunneling processes.  However it is not possible to obtain direct quasiparticle tunneling at a threshold bias of $e|V| = (\Delta_{Pb}+E_S)/2$, which can arise only through Andreev reflections where an electron from the Shiba state is reflected as a hole into the continuum. Fig.~\ref{fig:typeBshiba}(c)-(f) show that the spatial patterns of the Shiba states for the Type B adatom at a bias of $eV = \pm (\Delta_{Pb} + E_{S1})$ have a striking similarity to ones at the Andreev reflection process of $eV = \pm (\Delta_{Pb} + E_{S1})/2$, corroborating the role of the Shiba state in the Andreev process.  

\bibliography{supplementary_resub}{supplementary_resub}{}

\begin{thebibliography}{41}%
\makeatletter
\providecommand \@ifxundefined [1]{%
 \@ifx{#1\undefined}
}%
\providecommand \@ifnum [1]{%
 \ifnum #1\expandafter \@firstoftwo
 \else \expandafter \@secondoftwo
 \fi
}%
\providecommand \@ifx [1]{%
 \ifx #1\expandafter \@firstoftwo
 \else \expandafter \@secondoftwo
 \fi
}%
\providecommand \natexlab [1]{#1}%
\providecommand \enquote  [1]{``#1''}%
\providecommand \bibnamefont  [1]{#1}%
\providecommand \bibfnamefont [1]{#1}%
\providecommand \citenamefont [1]{#1}%
\providecommand \href@noop [0]{\@secondoftwo}%
\providecommand \href [0]{\begingroup \@sanitize@url \@href}%
\providecommand \@href[1]{\@@startlink{#1}\@@href}%
\providecommand \@@href[1]{\endgroup#1\@@endlink}%
\providecommand \@sanitize@url [0]{\catcode `\\12\catcode `\$12\catcode
  `\&12\catcode `\#12\catcode `\^12\catcode `\_12\catcode `\%12\relax}%
\providecommand \@@startlink[1]{}%
\providecommand \@@endlink[0]{}%
\providecommand \url  [0]{\begingroup\@sanitize@url \@url }%
\providecommand \@url [1]{\endgroup\@href {#1}{\urlprefix }}%
\providecommand \urlprefix  [0]{URL }%
\providecommand \Eprint [0]{\href }%
\providecommand \doibase [0]{http://dx.doi.org/}%
\providecommand \selectlanguage [0]{\@gobble}%
\providecommand \bibinfo  [0]{\@secondoftwo}%
\providecommand \bibfield  [0]{\@secondoftwo}%
\providecommand \translation [1]{[#1]}%
\providecommand \BibitemOpen [0]{}%
\providecommand \bibitemStop [0]{}%
\providecommand \bibitemNoStop [0]{.\EOS\space}%
\providecommand \EOS [0]{\spacefactor3000\relax}%
\providecommand \BibitemShut  [1]{\csname bibitem#1\endcsname}%
\let\auto@bib@innerbib\@empty
\bibitem [{\citenamefont {Sacepe}\ \emph {et~al.}(2011)\citenamefont {Sacepe},
  \citenamefont {Dubouchet}, \citenamefont {Chapelier}, \citenamefont
  {Sanquer}, \citenamefont {Ovadia}, \citenamefont {Shahar}, \citenamefont
  {Feigel'man},\ and\ \citenamefont {Ioffe}}]{sacepe2011}%
  \BibitemOpen
  \bibfield  {author} {\bibinfo {author} {\bibfnamefont {B.}~\bibnamefont
  {Sacepe}}, \bibinfo {author} {\bibfnamefont {T.}~\bibnamefont {Dubouchet}},
  \bibinfo {author} {\bibfnamefont {C.}~\bibnamefont {Chapelier}}, \bibinfo
  {author} {\bibfnamefont {M.}~\bibnamefont {Sanquer}}, \bibinfo {author}
  {\bibfnamefont {M.}~\bibnamefont {Ovadia}}, \bibinfo {author} {\bibfnamefont
  {D.}~\bibnamefont {Shahar}}, \bibinfo {author} {\bibfnamefont
  {M.}~\bibnamefont {Feigel'man}}, \ and\ \bibinfo {author} {\bibfnamefont
  {L.}~\bibnamefont {Ioffe}},\ }\href {\doibase 10.1038/nphys1892} {\bibfield
  {journal} {\bibinfo  {journal} {Nat. Phys.}\ }\textbf {\bibinfo {volume}
  {7}},\ \bibinfo {pages} {239} (\bibinfo {year} {2011})}\BibitemShut {NoStop}%
\bibitem [{\citenamefont {Bouadim}\ \emph {et~al.}(2011)\citenamefont
  {Bouadim}, \citenamefont {Loh}, \citenamefont {Randeria},\ and\ \citenamefont
  {Trivedi}}]{bouadim2011}%
  \BibitemOpen
  \bibfield  {author} {\bibinfo {author} {\bibfnamefont {K.}~\bibnamefont
  {Bouadim}}, \bibinfo {author} {\bibfnamefont {Y.~L.}\ \bibnamefont {Loh}},
  \bibinfo {author} {\bibfnamefont {M.}~\bibnamefont {Randeria}}, \ and\
  \bibinfo {author} {\bibfnamefont {N.}~\bibnamefont {Trivedi}},\ }\href
  {\doibase 10.1038/nphys2037} {\bibfield  {journal} {\bibinfo  {journal} {Nat.
  Phys.}\ }\textbf {\bibinfo {volume} {7}},\ \bibinfo {pages} {884} (\bibinfo
  {year} {2011})}\BibitemShut {NoStop}%
\bibitem [{\citenamefont {Gomes}\ \emph {et~al.}(2007)\citenamefont {Gomes},
  \citenamefont {Pasupathy}, \citenamefont {Pushp}, \citenamefont {Ono},
  \citenamefont {Ando},\ and\ \citenamefont {Yazdani}}]{gomes2007}%
  \BibitemOpen
  \bibfield  {author} {\bibinfo {author} {\bibfnamefont {K.~K.}\ \bibnamefont
  {Gomes}}, \bibinfo {author} {\bibfnamefont {A.~N.}\ \bibnamefont
  {Pasupathy}}, \bibinfo {author} {\bibfnamefont {A.}~\bibnamefont {Pushp}},
  \bibinfo {author} {\bibfnamefont {S.}~\bibnamefont {Ono}}, \bibinfo {author}
  {\bibfnamefont {Y.}~\bibnamefont {Ando}}, \ and\ \bibinfo {author}
  {\bibfnamefont {A.}~\bibnamefont {Yazdani}},\ }\href {\doibase
  10.1038/nature05881} {\bibfield  {journal} {\bibinfo  {journal} {Nature}\
  }\textbf {\bibinfo {volume} {447}},\ \bibinfo {pages} {569} (\bibinfo {year}
  {2007})}\BibitemShut {NoStop}%
\bibitem [{\citenamefont {Pasupathy}\ \emph {et~al.}(2008)\citenamefont
  {Pasupathy}, \citenamefont {Pushp}, \citenamefont {Gomes}, \citenamefont
  {Parker}, \citenamefont {Wen}, \citenamefont {Xu}, \citenamefont {Gu},
  \citenamefont {Ono}, \citenamefont {Ando},\ and\ \citenamefont
  {Yazdani}}]{pasupathy2008}%
  \BibitemOpen
  \bibfield  {author} {\bibinfo {author} {\bibfnamefont {A.~N.}\ \bibnamefont
  {Pasupathy}}, \bibinfo {author} {\bibfnamefont {A.}~\bibnamefont {Pushp}},
  \bibinfo {author} {\bibfnamefont {K.~K.}\ \bibnamefont {Gomes}}, \bibinfo
  {author} {\bibfnamefont {C.~V.}\ \bibnamefont {Parker}}, \bibinfo {author}
  {\bibfnamefont {J.}~\bibnamefont {Wen}}, \bibinfo {author} {\bibfnamefont
  {Z.}~\bibnamefont {Xu}}, \bibinfo {author} {\bibfnamefont {G.}~\bibnamefont
  {Gu}}, \bibinfo {author} {\bibfnamefont {S.}~\bibnamefont {Ono}}, \bibinfo
  {author} {\bibfnamefont {Y.}~\bibnamefont {Ando}}, \ and\ \bibinfo {author}
  {\bibfnamefont {A.}~\bibnamefont {Yazdani}},\ }\href {\doibase
  10.1126/science.1154700} {\bibfield  {journal} {\bibinfo  {journal}
  {Science}\ }\textbf {\bibinfo {volume} {320}},\ \bibinfo {pages} {196}
  (\bibinfo {year} {2008})}\BibitemShut {NoStop}%
\bibitem [{\citenamefont {Fulde}\ and\ \citenamefont
  {Ferrell}(1964)}]{fulde1964}%
  \BibitemOpen
  \bibfield  {author} {\bibinfo {author} {\bibfnamefont {P.}~\bibnamefont
  {Fulde}}\ and\ \bibinfo {author} {\bibfnamefont {R.~A.}\ \bibnamefont
  {Ferrell}},\ }\href {\doibase 10.1103/PhysRev.135.A550} {\bibfield  {journal}
  {\bibinfo  {journal} {Phys. Rev.}\ }\textbf {\bibinfo {volume} {135}},\
  \bibinfo {pages} {A550} (\bibinfo {year} {1964})}\BibitemShut {NoStop}%
\bibitem [{\citenamefont {Larkin}\ and\ \citenamefont
  {Ovchinnikov}(1965)}]{larkin1965}%
  \BibitemOpen
  \bibfield  {author} {\bibinfo {author} {\bibfnamefont {A.}~\bibnamefont
  {Larkin}}\ and\ \bibinfo {author} {\bibfnamefont {Y.}~\bibnamefont
  {Ovchinnikov}},\ }\href@noop {} {\bibfield  {journal} {\bibinfo  {journal}
  {Soviet Physics-JETP}\ }\textbf {\bibinfo {volume} {20}},\ \bibinfo {pages}
  {762} (\bibinfo {year} {1965})}\BibitemShut {NoStop}%
\bibitem [{\citenamefont {Matsuda}\ and\ \citenamefont
  {Shimahara}(2007)}]{matsuda2007}%
  \BibitemOpen
  \bibfield  {author} {\bibinfo {author} {\bibfnamefont {Y.}~\bibnamefont
  {Matsuda}}\ and\ \bibinfo {author} {\bibfnamefont {H.}~\bibnamefont
  {Shimahara}},\ }\href {\doibase 10.1143/JPSJ.76.051005} {\bibfield  {journal}
  {\bibinfo  {journal} {Journal of the Physical Society of Japan}\ }\textbf
  {\bibinfo {volume} {76}},\ \bibinfo {pages} {051005} (\bibinfo {year}
  {2007})}\BibitemShut {NoStop}%
\bibitem [{\citenamefont {Chen}\ \emph {et~al.}(2004)\citenamefont {Chen},
  \citenamefont {Vafek}, \citenamefont {Yazdani},\ and\ \citenamefont
  {Zhang}}]{chen2004}%
  \BibitemOpen
  \bibfield  {author} {\bibinfo {author} {\bibfnamefont {H.-D.}\ \bibnamefont
  {Chen}}, \bibinfo {author} {\bibfnamefont {O.}~\bibnamefont {Vafek}},
  \bibinfo {author} {\bibfnamefont {A.}~\bibnamefont {Yazdani}}, \ and\
  \bibinfo {author} {\bibfnamefont {S.-C.}\ \bibnamefont {Zhang}},\ }\href
  {\doibase 10.1103/PhysRevLett.93.187002} {\bibfield  {journal} {\bibinfo
  {journal} {Phys. Rev. Lett.}\ }\textbf {\bibinfo {volume} {93}},\ \bibinfo
  {pages} {187002} (\bibinfo {year} {2004})}\BibitemShut {NoStop}%
\bibitem [{\citenamefont {Berg}\ \emph {et~al.}(2009)\citenamefont {Berg},
  \citenamefont {Fradkin},\ and\ \citenamefont {Kivelson}}]{berg2009}%
  \BibitemOpen
  \bibfield  {author} {\bibinfo {author} {\bibfnamefont {E.}~\bibnamefont
  {Berg}}, \bibinfo {author} {\bibfnamefont {E.}~\bibnamefont {Fradkin}}, \
  and\ \bibinfo {author} {\bibfnamefont {S.~A.}\ \bibnamefont {Kivelson}},\
  }\href {\doibase 10.1038/nphys1389} {\bibfield  {journal} {\bibinfo
  {journal} {Nat. Phys.}\ }\textbf {\bibinfo {volume} {5}},\ \bibinfo {pages}
  {830} (\bibinfo {year} {2009})}\BibitemShut {NoStop}%
\bibitem [{\citenamefont {Lee}(2014)}]{lee2014}%
  \BibitemOpen
  \bibfield  {author} {\bibinfo {author} {\bibfnamefont {P.~A.}\ \bibnamefont
  {Lee}},\ }\href {\doibase 10.1103/PhysRevX.4.031017} {\bibfield  {journal}
  {\bibinfo  {journal} {Phys. Rev. X}\ }\textbf {\bibinfo {volume} {4}},\
  \bibinfo {pages} {031017} (\bibinfo {year} {2014})}\BibitemShut {NoStop}%
\bibitem [{\citenamefont {\ifmmode~\check{S}\else \v{S}\fi{}makov}\ \emph
  {et~al.}(2001)\citenamefont {\ifmmode~\check{S}\else \v{S}\fi{}makov},
  \citenamefont {Martin},\ and\ \citenamefont {Balatsky}}]{PhysRevB.64.212506}%
  \BibitemOpen
  \bibfield  {author} {\bibinfo {author} {\bibfnamefont {J.}~\bibnamefont
  {\ifmmode~\check{S}\else \v{S}\fi{}makov}}, \bibinfo {author} {\bibfnamefont
  {I.}~\bibnamefont {Martin}}, \ and\ \bibinfo {author} {\bibfnamefont {A.~V.}\
  \bibnamefont {Balatsky}},\ }\href {\doibase 10.1103/PhysRevB.64.212506}
  {\bibfield  {journal} {\bibinfo  {journal} {Phys. Rev. B}\ }\textbf {\bibinfo
  {volume} {64}},\ \bibinfo {pages} {212506} (\bibinfo {year}
  {2001})}\BibitemShut {NoStop}%
\bibitem [{\citenamefont {Rodrigo}\ and\ \citenamefont
  {Vieira}(2004)}]{rodrigo2004}%
  \BibitemOpen
  \bibfield  {author} {\bibinfo {author} {\bibfnamefont {J.}~\bibnamefont
  {Rodrigo}}\ and\ \bibinfo {author} {\bibfnamefont {S.}~\bibnamefont
  {Vieira}},\ }\href {\doibase http://dx.doi.org/10.1016/j.physc.2003.10.030}
  {\bibfield  {journal} {\bibinfo  {journal} {Physica C: Superconductivity}\
  }\textbf {\bibinfo {volume} {404}},\ \bibinfo {pages} {306 } (\bibinfo {year}
  {2004})}\BibitemShut {NoStop}%
\bibitem [{\citenamefont {Naaman}\ \emph {et~al.}(2001)\citenamefont {Naaman},
  \citenamefont {Teizer},\ and\ \citenamefont {Dynes}}]{PhysRevLett.87.097004}%
  \BibitemOpen
  \bibfield  {author} {\bibinfo {author} {\bibfnamefont {O.}~\bibnamefont
  {Naaman}}, \bibinfo {author} {\bibfnamefont {W.}~\bibnamefont {Teizer}}, \
  and\ \bibinfo {author} {\bibfnamefont {R.~C.}\ \bibnamefont {Dynes}},\ }\href
  {\doibase 10.1103/PhysRevLett.87.097004} {\bibfield  {journal} {\bibinfo
  {journal} {Phys. Rev. Lett.}\ }\textbf {\bibinfo {volume} {87}},\ \bibinfo
  {pages} {097004} (\bibinfo {year} {2001})}\BibitemShut {NoStop}%
\bibitem [{\citenamefont {{Rodrigo, J. G.}}\ \emph {et~al.}(2004)\citenamefont
  {{Rodrigo, J. G.}}, \citenamefont {{Suderow, H.}},\ and\ \citenamefont
  {{Vieira, S.}}}]{rodrigoEuroPHys}%
  \BibitemOpen
  \bibfield  {author} {\bibinfo {author} {\bibnamefont {{Rodrigo, J. G.}}},
  \bibinfo {author} {\bibnamefont {{Suderow, H.}}}, \ and\ \bibinfo {author}
  {\bibnamefont {{Vieira, S.}}},\ }\href {\doibase 10.1140/epjb/e2004-00273-y}
  {\bibfield  {journal} {\bibinfo  {journal} {Eur. Phys. J. B}\ }\textbf
  {\bibinfo {volume} {40}},\ \bibinfo {pages} {483} (\bibinfo {year}
  {2004})}\BibitemShut {NoStop}%
\bibitem [{\citenamefont {Levy}\ \emph {et~al.}(2013)\citenamefont {Levy},
  \citenamefont {Zhang}, \citenamefont {Ha}, \citenamefont {Sharifi},
  \citenamefont {Talin}, \citenamefont {Kuk},\ and\ \citenamefont
  {Stroscio}}]{levy2013}%
  \BibitemOpen
  \bibfield  {author} {\bibinfo {author} {\bibfnamefont {N.}~\bibnamefont
  {Levy}}, \bibinfo {author} {\bibfnamefont {T.}~\bibnamefont {Zhang}},
  \bibinfo {author} {\bibfnamefont {J.}~\bibnamefont {Ha}}, \bibinfo {author}
  {\bibfnamefont {F.}~\bibnamefont {Sharifi}}, \bibinfo {author} {\bibfnamefont
  {A.~A.}\ \bibnamefont {Talin}}, \bibinfo {author} {\bibfnamefont
  {Y.}~\bibnamefont {Kuk}}, \ and\ \bibinfo {author} {\bibfnamefont {J.~A.}\
  \bibnamefont {Stroscio}},\ }\href {\doibase 10.1103/PhysRevLett.110.117001}
  {\bibfield  {journal} {\bibinfo  {journal} {Phys. Rev. Lett.}\ }\textbf
  {\bibinfo {volume} {110}},\ \bibinfo {pages} {117001} (\bibinfo {year}
  {2013})}\BibitemShut {NoStop}%
\bibitem [{\citenamefont {Proslier}\ \emph {et~al.}(2006)\citenamefont
  {Proslier}, \citenamefont {Kohen}, \citenamefont {Noat}, \citenamefont
  {Cren}, \citenamefont {Roditchev},\ and\ \citenamefont
  {Sacks}}]{proslier2006}%
  \BibitemOpen
  \bibfield  {author} {\bibinfo {author} {\bibfnamefont {T.}~\bibnamefont
  {Proslier}}, \bibinfo {author} {\bibfnamefont {A.}~\bibnamefont {Kohen}},
  \bibinfo {author} {\bibfnamefont {Y.}~\bibnamefont {Noat}}, \bibinfo {author}
  {\bibfnamefont {T.}~\bibnamefont {Cren}}, \bibinfo {author} {\bibfnamefont
  {D.}~\bibnamefont {Roditchev}}, \ and\ \bibinfo {author} {\bibfnamefont
  {W.}~\bibnamefont {Sacks}},\ }\href
  {http://stacks.iop.org/0295-5075/73/i=6/a=962} {\bibfield  {journal}
  {\bibinfo  {journal} {Europhys. Lett.}\ }\textbf {\bibinfo {volume} {73}},\
  \bibinfo {pages} {962} (\bibinfo {year} {2006})}\BibitemShut {NoStop}%
\bibitem [{\citenamefont {Bergeal}\ \emph {et~al.}(2008)\citenamefont
  {Bergeal}, \citenamefont {Noat}, \citenamefont {Cren}, \citenamefont
  {Proslier}, \citenamefont {Dubost}, \citenamefont {Debontridder},
  \citenamefont {Zimmers}, \citenamefont {Roditchev}, \citenamefont {Sacks},\
  and\ \citenamefont {Marcus}}]{PhysRevB.78.140507}%
  \BibitemOpen
  \bibfield  {author} {\bibinfo {author} {\bibfnamefont {N.}~\bibnamefont
  {Bergeal}}, \bibinfo {author} {\bibfnamefont {Y.}~\bibnamefont {Noat}},
  \bibinfo {author} {\bibfnamefont {T.}~\bibnamefont {Cren}}, \bibinfo {author}
  {\bibfnamefont {T.}~\bibnamefont {Proslier}}, \bibinfo {author}
  {\bibfnamefont {V.}~\bibnamefont {Dubost}}, \bibinfo {author} {\bibfnamefont
  {F.}~\bibnamefont {Debontridder}}, \bibinfo {author} {\bibfnamefont
  {A.}~\bibnamefont {Zimmers}}, \bibinfo {author} {\bibfnamefont
  {D.}~\bibnamefont {Roditchev}}, \bibinfo {author} {\bibfnamefont
  {W.}~\bibnamefont {Sacks}}, \ and\ \bibinfo {author} {\bibfnamefont
  {J.}~\bibnamefont {Marcus}},\ }\href {\doibase 10.1103/PhysRevB.78.140507}
  {\bibfield  {journal} {\bibinfo  {journal} {Phys. Rev. B}\ }\textbf {\bibinfo
  {volume} {78}},\ \bibinfo {pages} {140507} (\bibinfo {year}
  {2008})}\BibitemShut {NoStop}%
\bibitem [{\citenamefont {Kimura}\ \emph {et~al.}(2009)\citenamefont {Kimura},
  \citenamefont {Barber}, \citenamefont {Ono}, \citenamefont {Ando},\ and\
  \citenamefont {Dynes}}]{PhysRevB.80.144506}%
  \BibitemOpen
  \bibfield  {author} {\bibinfo {author} {\bibfnamefont {H.}~\bibnamefont
  {Kimura}}, \bibinfo {author} {\bibfnamefont {R.~P.}\ \bibnamefont {Barber}},
  \bibinfo {author} {\bibfnamefont {S.}~\bibnamefont {Ono}}, \bibinfo {author}
  {\bibfnamefont {Y.}~\bibnamefont {Ando}}, \ and\ \bibinfo {author}
  {\bibfnamefont {R.~C.}\ \bibnamefont {Dynes}},\ }\href {\doibase
  10.1103/PhysRevB.80.144506} {\bibfield  {journal} {\bibinfo  {journal} {Phys.
  Rev. B}\ }\textbf {\bibinfo {volume} {80}},\ \bibinfo {pages} {144506}
  (\bibinfo {year} {2009})}\BibitemShut {NoStop}%
\bibitem [{\citenamefont {Hamidian}\ \emph {et~al.}(2015)\citenamefont
  {Hamidian}, \citenamefont {Edkins}, \citenamefont {Joo}, \citenamefont
  {Kostin}, \citenamefont {Eisaki}, \citenamefont {Uchida}, \citenamefont
  {Lawler}, \citenamefont {Kim}, \citenamefont {Mackenzie}, \citenamefont
  {Fujita}, \citenamefont {Lee},\ and\ \citenamefont {Davis}}]{davis2015}%
  \BibitemOpen
  \bibfield  {author} {\bibinfo {author} {\bibfnamefont {M.~H.}\ \bibnamefont
  {Hamidian}}, \bibinfo {author} {\bibfnamefont {S.~D.}\ \bibnamefont
  {Edkins}}, \bibinfo {author} {\bibfnamefont {S.~H.}\ \bibnamefont {Joo}},
  \bibinfo {author} {\bibfnamefont {A.}~\bibnamefont {Kostin}}, \bibinfo
  {author} {\bibfnamefont {H.}~\bibnamefont {Eisaki}}, \bibinfo {author}
  {\bibfnamefont {S.}~\bibnamefont {Uchida}}, \bibinfo {author} {\bibfnamefont
  {M.~J.}\ \bibnamefont {Lawler}}, \bibinfo {author} {\bibfnamefont {E.~A.}\
  \bibnamefont {Kim}}, \bibinfo {author} {\bibfnamefont {A.~P.}\ \bibnamefont
  {Mackenzie}}, \bibinfo {author} {\bibfnamefont {K.}~\bibnamefont {Fujita}},
  \bibinfo {author} {\bibfnamefont {J.}~\bibnamefont {Lee}}, \ and\ \bibinfo
  {author} {\bibfnamefont {S.~J.~C.}\ \bibnamefont {Davis}},\ }\href@noop {} {\
   (\bibinfo {year} {2015})},\ \Eprint {http://arxiv.org/abs/1511.08124}
  {arXiv:1511.08124} \BibitemShut {NoStop}%
\bibitem [{\citenamefont {Flatt\'e}\ and\ \citenamefont
  {Byers}(1997{\natexlab{a}})}]{FlattePRL}%
  \BibitemOpen
  \bibfield  {author} {\bibinfo {author} {\bibfnamefont {M.~E.}\ \bibnamefont
  {Flatt\'e}}\ and\ \bibinfo {author} {\bibfnamefont {J.~M.}\ \bibnamefont
  {Byers}},\ }\href {\doibase 10.1103/PhysRevLett.78.3761} {\bibfield
  {journal} {\bibinfo  {journal} {Phys. Rev. Lett.}\ }\textbf {\bibinfo
  {volume} {78}},\ \bibinfo {pages} {3761} (\bibinfo {year}
  {1997}{\natexlab{a}})}\BibitemShut {NoStop}%
\bibitem [{\citenamefont {Flatt\'e}\ and\ \citenamefont
  {Byers}(1997{\natexlab{b}})}]{FlattePRB}%
  \BibitemOpen
  \bibfield  {author} {\bibinfo {author} {\bibfnamefont {M.~E.}\ \bibnamefont
  {Flatt\'e}}\ and\ \bibinfo {author} {\bibfnamefont {J.~M.}\ \bibnamefont
  {Byers}},\ }\href {\doibase 10.1103/PhysRevB.56.11213} {\bibfield  {journal}
  {\bibinfo  {journal} {Phys. Rev. B}\ }\textbf {\bibinfo {volume} {56}},\
  \bibinfo {pages} {11213} (\bibinfo {year} {1997}{\natexlab{b}})}\BibitemShut
  {NoStop}%
\bibitem [{\citenamefont {Misra}\ \emph {et~al.}(2013)\citenamefont {Misra},
  \citenamefont {Zhou}, \citenamefont {Drozdov}, \citenamefont {Seo},
  \citenamefont {Urban}, \citenamefont {Gyenis}, \citenamefont {Kingsley},
  \citenamefont {Jones},\ and\ \citenamefont {Yazdani}}]{misra2013}%
  \BibitemOpen
  \bibfield  {author} {\bibinfo {author} {\bibfnamefont {S.}~\bibnamefont
  {Misra}}, \bibinfo {author} {\bibfnamefont {B.~B.}\ \bibnamefont {Zhou}},
  \bibinfo {author} {\bibfnamefont {I.~K.}\ \bibnamefont {Drozdov}}, \bibinfo
  {author} {\bibfnamefont {J.}~\bibnamefont {Seo}}, \bibinfo {author}
  {\bibfnamefont {L.}~\bibnamefont {Urban}}, \bibinfo {author} {\bibfnamefont
  {A.}~\bibnamefont {Gyenis}}, \bibinfo {author} {\bibfnamefont {S.~C.~J.}\
  \bibnamefont {Kingsley}}, \bibinfo {author} {\bibfnamefont {H.}~\bibnamefont
  {Jones}}, \ and\ \bibinfo {author} {\bibfnamefont {A.}~\bibnamefont
  {Yazdani}},\ }\href {\doibase http://dx.doi.org/10.1063/1.4822271} {\bibfield
   {journal} {\bibinfo  {journal} {Rev. Sci. Inst.}\ }\textbf {\bibinfo
  {volume} {84}},\ \bibinfo {eid} {103903} (\bibinfo {year} {2013}),\
  http://dx.doi.org/10.1063/1.4822271}\BibitemShut {NoStop}%
\bibitem [{\citenamefont {Townsend}\ and\ \citenamefont
  {Sutton}(1962)}]{PhysRev.128.591}%
  \BibitemOpen
  \bibfield  {author} {\bibinfo {author} {\bibfnamefont {P.}~\bibnamefont
  {Townsend}}\ and\ \bibinfo {author} {\bibfnamefont {J.}~\bibnamefont
  {Sutton}},\ }\href {\doibase 10.1103/PhysRev.128.591} {\bibfield  {journal}
  {\bibinfo  {journal} {Phys. Rev.}\ }\textbf {\bibinfo {volume} {128}},\
  \bibinfo {pages} {591} (\bibinfo {year} {1962})}\BibitemShut {NoStop}%
\bibitem [{\citenamefont {Blackford}\ and\ \citenamefont
  {March}(1969)}]{PhysRev.186.397}%
  \BibitemOpen
  \bibfield  {author} {\bibinfo {author} {\bibfnamefont {B.~L.}\ \bibnamefont
  {Blackford}}\ and\ \bibinfo {author} {\bibfnamefont {R.~H.}\ \bibnamefont
  {March}},\ }\href {\doibase 10.1103/PhysRev.186.397} {\bibfield  {journal}
  {\bibinfo  {journal} {Phys. Rev.}\ }\textbf {\bibinfo {volume} {186}},\
  \bibinfo {pages} {397} (\bibinfo {year} {1969})}\BibitemShut {NoStop}%
\bibitem [{\citenamefont {Ruby}\ \emph
  {et~al.}(2015{\natexlab{a}})\citenamefont {Ruby}, \citenamefont {Heinrich},
  \citenamefont {Pascual},\ and\ \citenamefont {Franke}}]{ruby2015}%
  \BibitemOpen
  \bibfield  {author} {\bibinfo {author} {\bibfnamefont {M.}~\bibnamefont
  {Ruby}}, \bibinfo {author} {\bibfnamefont {B.~W.}\ \bibnamefont {Heinrich}},
  \bibinfo {author} {\bibfnamefont {J.~I.}\ \bibnamefont {Pascual}}, \ and\
  \bibinfo {author} {\bibfnamefont {K.~J.}\ \bibnamefont {Franke}},\ }\href
  {\doibase 10.1103/PhysRevLett.114.157001} {\bibfield  {journal} {\bibinfo
  {journal} {Phys. Rev. Lett.}\ }\textbf {\bibinfo {volume} {114}},\ \bibinfo
  {pages} {157001} (\bibinfo {year} {2015}{\natexlab{a}})}\BibitemShut
  {NoStop}%
\bibitem [{\citenamefont {Ternes}\ \emph {et~al.}(2006)\citenamefont {Ternes},
  \citenamefont {Schneider}, \citenamefont {Cuevas}, \citenamefont {Lutz},
  \citenamefont {Hirjibehedin},\ and\ \citenamefont {Heinrich}}]{heinrich2006}%
  \BibitemOpen
  \bibfield  {author} {\bibinfo {author} {\bibfnamefont {M.}~\bibnamefont
  {Ternes}}, \bibinfo {author} {\bibfnamefont {W.-D.}\ \bibnamefont
  {Schneider}}, \bibinfo {author} {\bibfnamefont {J.-C.}\ \bibnamefont
  {Cuevas}}, \bibinfo {author} {\bibfnamefont {C.~P.}\ \bibnamefont {Lutz}},
  \bibinfo {author} {\bibfnamefont {C.~F.}\ \bibnamefont {Hirjibehedin}}, \
  and\ \bibinfo {author} {\bibfnamefont {A.~J.}\ \bibnamefont {Heinrich}},\
  }\href {\doibase 10.1103/PhysRevB.74.132501} {\bibfield  {journal} {\bibinfo
  {journal} {Phys. Rev. B}\ }\textbf {\bibinfo {volume} {74}},\ \bibinfo
  {pages} {132501} (\bibinfo {year} {2006})}\BibitemShut {NoStop}%
\bibitem [{\citenamefont {Ingold}\ \emph {et~al.}(1994)\citenamefont {Ingold},
  \citenamefont {Grabert},\ and\ \citenamefont {Eberhardt}}]{PhysRevB.50.395}%
  \BibitemOpen
  \bibfield  {author} {\bibinfo {author} {\bibfnamefont {G.-L.}\ \bibnamefont
  {Ingold}}, \bibinfo {author} {\bibfnamefont {H.}~\bibnamefont {Grabert}}, \
  and\ \bibinfo {author} {\bibfnamefont {U.}~\bibnamefont {Eberhardt}},\ }\href
  {\doibase 10.1103/PhysRevB.50.395} {\bibfield  {journal} {\bibinfo  {journal}
  {Phys. Rev. B}\ }\textbf {\bibinfo {volume} {50}},\ \bibinfo {pages} {395}
  (\bibinfo {year} {1994})}\BibitemShut {NoStop}%
\bibitem [{\citenamefont {Jaeck}\ \emph {et~al.}(2015)\citenamefont {Jaeck},
  \citenamefont {Eltschka}, \citenamefont {Assig}, \citenamefont {Hardock},
  \citenamefont {Etzkorn}, \citenamefont {Ast},\ and\ \citenamefont
  {Kern}}]{jack2015}%
  \BibitemOpen
  \bibfield  {author} {\bibinfo {author} {\bibfnamefont {B.}~\bibnamefont
  {Jaeck}}, \bibinfo {author} {\bibfnamefont {M.}~\bibnamefont {Eltschka}},
  \bibinfo {author} {\bibfnamefont {M.}~\bibnamefont {Assig}}, \bibinfo
  {author} {\bibfnamefont {A.}~\bibnamefont {Hardock}}, \bibinfo {author}
  {\bibfnamefont {M.}~\bibnamefont {Etzkorn}}, \bibinfo {author} {\bibfnamefont
  {C.~R.}\ \bibnamefont {Ast}}, \ and\ \bibinfo {author} {\bibfnamefont
  {K.}~\bibnamefont {Kern}},\ }\href {\doibase
  http://dx.doi.org/10.1063/1.4905322} {\bibfield  {journal} {\bibinfo
  {journal} {App. Phys. Lett.}\ }\textbf {\bibinfo {volume} {106}},\ \bibinfo
  {eid} {013109} (\bibinfo {year} {2015}),\
  http://dx.doi.org/10.1063/1.4905322}\BibitemShut {NoStop}%
\bibitem [{\citenamefont {Roychowdhury}\ \emph {et~al.}(2015)\citenamefont
  {Roychowdhury}, \citenamefont {Dreyer}, \citenamefont {Anderson},
  \citenamefont {Lobb},\ and\ \citenamefont {Wellstood}}]{roychowdhury2015}%
  \BibitemOpen
  \bibfield  {author} {\bibinfo {author} {\bibfnamefont {A.}~\bibnamefont
  {Roychowdhury}}, \bibinfo {author} {\bibfnamefont {M.}~\bibnamefont
  {Dreyer}}, \bibinfo {author} {\bibfnamefont {J.~R.}\ \bibnamefont
  {Anderson}}, \bibinfo {author} {\bibfnamefont {C.~J.}\ \bibnamefont {Lobb}},
  \ and\ \bibinfo {author} {\bibfnamefont {F.~C.}\ \bibnamefont {Wellstood}},\
  }\href {\doibase 10.1103/PhysRevApplied.4.034011} {\bibfield  {journal}
  {\bibinfo  {journal} {Phys. Rev. Applied}\ }\textbf {\bibinfo {volume} {4}},\
  \bibinfo {pages} {034011} (\bibinfo {year} {2015})}\BibitemShut {NoStop}%
\bibitem [{\citenamefont {Ingold}\ and\ \citenamefont
  {Grabert}(1991)}]{ingoldGrabert1991}%
  \BibitemOpen
  \bibfield  {author} {\bibinfo {author} {\bibfnamefont {G.-L.}\ \bibnamefont
  {Ingold}}\ and\ \bibinfo {author} {\bibfnamefont {H.}~\bibnamefont
  {Grabert}},\ }\href {http://stacks.iop.org/0295-5075/14/i=4/a=015} {\bibfield
   {journal} {\bibinfo  {journal} {EPL (Europhysics Letters)}\ }\textbf
  {\bibinfo {volume} {14}},\ \bibinfo {pages} {371} (\bibinfo {year}
  {1991})}\BibitemShut {NoStop}%
\bibitem [{\citenamefont {Ambegaokar}\ and\ \citenamefont
  {Baratoff}(1963)}]{ambegaokar1963}%
  \BibitemOpen
  \bibfield  {author} {\bibinfo {author} {\bibfnamefont {V.}~\bibnamefont
  {Ambegaokar}}\ and\ \bibinfo {author} {\bibfnamefont {A.}~\bibnamefont
  {Baratoff}},\ }\href {\doibase 10.1103/PhysRevLett.10.486} {\bibfield
  {journal} {\bibinfo  {journal} {Phys. Rev. Lett.}\ }\textbf {\bibinfo
  {volume} {10}},\ \bibinfo {pages} {486} (\bibinfo {year} {1963})}\BibitemShut
  {NoStop}%
\bibitem [{\citenamefont {Anderson}(1959)}]{anderson1959}%
  \BibitemOpen
  \bibfield  {author} {\bibinfo {author} {\bibfnamefont {P.}~\bibnamefont
  {Anderson}},\ }\href {\doibase
  http://dx.doi.org/10.1016/0022-3697(59)90036-8} {\bibfield  {journal}
  {\bibinfo  {journal} {Journal of Physics and Chemistry of Solids}\ }\textbf
  {\bibinfo {volume} {11}},\ \bibinfo {pages} {26 } (\bibinfo {year}
  {1959})}\BibitemShut {NoStop}%
\bibitem [{\citenamefont {Abrikosov}\ and\ \citenamefont
  {Gor'kov}(1961)}]{abrikosov1961}%
  \BibitemOpen
  \bibfield  {author} {\bibinfo {author} {\bibfnamefont {A.~A.}\ \bibnamefont
  {Abrikosov}}\ and\ \bibinfo {author} {\bibfnamefont {L.~P.}\ \bibnamefont
  {Gor'kov}},\ }\href@noop {} {\bibfield  {journal} {\bibinfo  {journal} {Sov.
  Phys. JETP}\ }\textbf {\bibinfo {volume} {12}},\ \bibinfo {pages} {1243}
  (\bibinfo {year} {1961})}\BibitemShut {NoStop}%
\bibitem [{\citenamefont {Shiba}(1968)}]{shiba1968}%
  \BibitemOpen
  \bibfield  {author} {\bibinfo {author} {\bibfnamefont {H.}~\bibnamefont
  {Shiba}},\ }\href@noop {} {\bibfield  {journal} {\bibinfo  {journal} {Prog.
  Theor. Phys.}\ }\textbf {\bibinfo {volume} {40}},\ \bibinfo {pages} {435}
  (\bibinfo {year} {1968})}\BibitemShut {NoStop}%
\bibitem [{\citenamefont {Yu}(1965)}]{yu1965}%
  \BibitemOpen
  \bibfield  {author} {\bibinfo {author} {\bibfnamefont {L.}~\bibnamefont
  {Yu}},\ }\href@noop {} {\bibfield  {journal} {\bibinfo  {journal} {Acta Phys.
  Sin.}\ }\textbf {\bibinfo {volume} {21}},\ \bibinfo {pages} {75} (\bibinfo
  {year} {1965})}\BibitemShut {NoStop}%
\bibitem [{\citenamefont {Rusinov}(1969)}]{rusinov1969}%
  \BibitemOpen
  \bibfield  {author} {\bibinfo {author} {\bibfnamefont {A.~I.}\ \bibnamefont
  {Rusinov}},\ }\href@noop {} {\bibfield  {journal} {\bibinfo  {journal} {Sov.
  Phys. JETP}\ }\textbf {\bibinfo {volume} {9}},\ \bibinfo {pages} {85}
  (\bibinfo {year} {1969})}\BibitemShut {NoStop}%
\bibitem [{\citenamefont {Yazdani}\ \emph {et~al.}(1997)\citenamefont
  {Yazdani}, \citenamefont {Jones}, \citenamefont {Lutz}, \citenamefont
  {Crommie},\ and\ \citenamefont {Eigler}}]{yazdani1997}%
  \BibitemOpen
  \bibfield  {author} {\bibinfo {author} {\bibfnamefont {A.}~\bibnamefont
  {Yazdani}}, \bibinfo {author} {\bibfnamefont {B.~A.}\ \bibnamefont {Jones}},
  \bibinfo {author} {\bibfnamefont {C.~P.}\ \bibnamefont {Lutz}}, \bibinfo
  {author} {\bibfnamefont {M.~F.}\ \bibnamefont {Crommie}}, \ and\ \bibinfo
  {author} {\bibfnamefont {D.~M.}\ \bibnamefont {Eigler}},\ }\href {\doibase
  10.1126/science.275.5307.1767} {\bibfield  {journal} {\bibinfo  {journal}
  {Science}\ }\textbf {\bibinfo {volume} {275}},\ \bibinfo {pages} {1767}
  (\bibinfo {year} {1997})}\BibitemShut {NoStop}%
\bibitem [{\citenamefont {Ji}\ \emph {et~al.}(2008)\citenamefont {Ji},
  \citenamefont {Zhang}, \citenamefont {Fu}, \citenamefont {Chen},
  \citenamefont {Ma}, \citenamefont {Li}, \citenamefont {Duan}, \citenamefont
  {Jia},\ and\ \citenamefont {Xue}}]{PhysRevLett.100.226801}%
  \BibitemOpen
  \bibfield  {author} {\bibinfo {author} {\bibfnamefont {S.-H.}\ \bibnamefont
  {Ji}}, \bibinfo {author} {\bibfnamefont {T.}~\bibnamefont {Zhang}}, \bibinfo
  {author} {\bibfnamefont {Y.-S.}\ \bibnamefont {Fu}}, \bibinfo {author}
  {\bibfnamefont {X.}~\bibnamefont {Chen}}, \bibinfo {author} {\bibfnamefont
  {X.-C.}\ \bibnamefont {Ma}}, \bibinfo {author} {\bibfnamefont
  {J.}~\bibnamefont {Li}}, \bibinfo {author} {\bibfnamefont {W.-H.}\
  \bibnamefont {Duan}}, \bibinfo {author} {\bibfnamefont {J.-F.}\ \bibnamefont
  {Jia}}, \ and\ \bibinfo {author} {\bibfnamefont {Q.-K.}\ \bibnamefont
  {Xue}},\ }\href {\doibase 10.1103/PhysRevLett.100.226801} {\bibfield
  {journal} {\bibinfo  {journal} {Phys. Rev. Lett.}\ }\textbf {\bibinfo
  {volume} {100}},\ \bibinfo {pages} {226801} (\bibinfo {year}
  {2008})}\BibitemShut {NoStop}%
\bibitem [{\citenamefont {Heinrichs}(1968)}]{heinrichs1968}%
  \BibitemOpen
  \bibfield  {author} {\bibinfo {author} {\bibfnamefont {J.}~\bibnamefont
  {Heinrichs}},\ }\href {\doibase 10.1103/PhysRev.168.451} {\bibfield
  {journal} {\bibinfo  {journal} {Phys. Rev.}\ }\textbf {\bibinfo {volume}
  {168}},\ \bibinfo {pages} {451} (\bibinfo {year} {1968})}\BibitemShut
  {NoStop}%
\bibitem [{\citenamefont {Schlottmann}(1976)}]{schlottmann1976}%
  \BibitemOpen
  \bibfield  {author} {\bibinfo {author} {\bibfnamefont {P.}~\bibnamefont
  {Schlottmann}},\ }\href {\doibase 10.1103/PhysRevB.13.1} {\bibfield
  {journal} {\bibinfo  {journal} {Phys. Rev. B}\ }\textbf {\bibinfo {volume}
  {13}},\ \bibinfo {pages} {1} (\bibinfo {year} {1976})}\BibitemShut {NoStop}%
\bibitem [{\citenamefont {Ruby}\ \emph
  {et~al.}(2015{\natexlab{b}})\citenamefont {Ruby}, \citenamefont {Pientka},
  \citenamefont {Peng}, \citenamefont {von Oppen}, \citenamefont {Heinrich},\
  and\ \citenamefont {Franke}}]{PhysRevLett.115.087001}%
  \BibitemOpen
  \bibfield  {author} {\bibinfo {author} {\bibfnamefont {M.}~\bibnamefont
  {Ruby}}, \bibinfo {author} {\bibfnamefont {F.}~\bibnamefont {Pientka}},
  \bibinfo {author} {\bibfnamefont {Y.}~\bibnamefont {Peng}}, \bibinfo {author}
  {\bibfnamefont {F.}~\bibnamefont {von Oppen}}, \bibinfo {author}
  {\bibfnamefont {B.~W.}\ \bibnamefont {Heinrich}}, \ and\ \bibinfo {author}
  {\bibfnamefont {K.~J.}\ \bibnamefont {Franke}},\ }\href {\doibase
  10.1103/PhysRevLett.115.087001} {\bibfield  {journal} {\bibinfo  {journal}
  {Phys. Rev. Lett.}\ }\textbf {\bibinfo {volume} {115}},\ \bibinfo {pages}
  {087001} (\bibinfo {year} {2015}{\natexlab{b}})}\BibitemShut {NoStop}%
\end{thebibliography}%


\begin{thebibliography}{6}%
\makeatletter
\providecommand \@ifxundefined [1]{%
 \@ifx{#1\undefined}
}%
\providecommand \@ifnum [1]{%
 \ifnum #1\expandafter \@firstoftwo
 \else \expandafter \@secondoftwo
 \fi
}%
\providecommand \@ifx [1]{%
 \ifx #1\expandafter \@firstoftwo
 \else \expandafter \@secondoftwo
 \fi
}%
\providecommand \natexlab [1]{#1}%
\providecommand \enquote  [1]{``#1''}%
\providecommand \bibnamefont  [1]{#1}%
\providecommand \bibfnamefont [1]{#1}%
\providecommand \citenamefont [1]{#1}%
\providecommand \href@noop [0]{\@secondoftwo}%
\providecommand \href [0]{\begingroup \@sanitize@url \@href}%
\providecommand \@href[1]{\@@startlink{#1}\@@href}%
\providecommand \@@href[1]{\endgroup#1\@@endlink}%
\providecommand \@sanitize@url [0]{\catcode `\\12\catcode `\$12\catcode
  `\&12\catcode `\#12\catcode `\^12\catcode `\_12\catcode `\%12\relax}%
\providecommand \@@startlink[1]{}%
\providecommand \@@endlink[0]{}%
\providecommand \url  [0]{\begingroup\@sanitize@url \@url }%
\providecommand \@url [1]{\endgroup\@href {#1}{\urlprefix }}%
\providecommand \urlprefix  [0]{URL }%
\providecommand \Eprint [0]{\href }%
\providecommand \doibase [0]{http://dx.doi.org/}%
\providecommand \selectlanguage [0]{\@gobble}%
\providecommand \bibinfo  [0]{\@secondoftwo}%
\providecommand \bibfield  [0]{\@secondoftwo}%
\providecommand \translation [1]{[#1]}%
\providecommand \BibitemOpen [0]{}%
\providecommand \bibitemStop [0]{}%
\providecommand \bibitemNoStop [0]{.\EOS\space}%
\providecommand \EOS [0]{\spacefactor3000\relax}%
\providecommand \BibitemShut  [1]{\csname bibitem#1\endcsname}%
\let\auto@bib@innerbib\@empty
\bibitem [{\citenamefont {Devoret}\ \emph {et~al.}(1990)\citenamefont
  {Devoret}, \citenamefont {Esteve}, \citenamefont {Grabert}, \citenamefont
  {Ingold}, \citenamefont {Pothier},\ and\ \citenamefont
  {Urbina}}]{supp_PhysRevLett.64.1824}%
  \BibitemOpen
  \bibfield  {author} {\bibinfo {author} {\bibfnamefont {M.~H.}\ \bibnamefont
  {Devoret}}, \bibinfo {author} {\bibfnamefont {D.}~\bibnamefont {Esteve}},
  \bibinfo {author} {\bibfnamefont {H.}~\bibnamefont {Grabert}}, \bibinfo
  {author} {\bibfnamefont {G.-L.}\ \bibnamefont {Ingold}}, \bibinfo {author}
  {\bibfnamefont {H.}~\bibnamefont {Pothier}}, \ and\ \bibinfo {author}
  {\bibfnamefont {C.}~\bibnamefont {Urbina}},\ }\href {\doibase
  10.1103/PhysRevLett.64.1824} {\bibfield  {journal} {\bibinfo  {journal}
  {Phys. Rev. Lett.}\ }\textbf {\bibinfo {volume} {64}},\ \bibinfo {pages}
  {1824} (\bibinfo {year} {1990})}\BibitemShut {NoStop}%
\bibitem [{\citenamefont {Ingold}\ and\ \citenamefont
  {Grabert}(1991)}]{supp_ingoldGrabert1991}%
  \BibitemOpen
  \bibfield  {author} {\bibinfo {author} {\bibfnamefont {G.-L.}\ \bibnamefont
  {Ingold}}\ and\ \bibinfo {author} {\bibfnamefont {H.}~\bibnamefont
  {Grabert}},\ }\href {http://stacks.iop.org/0295-5075/14/i=4/a=015} {\bibfield
   {journal} {\bibinfo  {journal} {EPL (Europhysics Letters)}\ }\textbf
  {\bibinfo {volume} {14}},\ \bibinfo {pages} {371} (\bibinfo {year}
  {1991})}\BibitemShut {NoStop}%
\bibitem [{\citenamefont {Ingold}\ \emph {et~al.}(1994)\citenamefont {Ingold},
  \citenamefont {Grabert},\ and\ \citenamefont
  {Eberhardt}}]{supp_PhysRevB.50.395}%
  \BibitemOpen
  \bibfield  {author} {\bibinfo {author} {\bibfnamefont {G.-L.}\ \bibnamefont
  {Ingold}}, \bibinfo {author} {\bibfnamefont {H.}~\bibnamefont {Grabert}}, \
  and\ \bibinfo {author} {\bibfnamefont {U.}~\bibnamefont {Eberhardt}},\ }\href
  {\doibase 10.1103/PhysRevB.50.395} {\bibfield  {journal} {\bibinfo  {journal}
  {Phys. Rev. B}\ }\textbf {\bibinfo {volume} {50}},\ \bibinfo {pages} {395}
  (\bibinfo {year} {1994})}\BibitemShut {NoStop}%
\bibitem [{\citenamefont {Jaeck}\ \emph {et~al.}(2015)\citenamefont {Jaeck},
  \citenamefont {Eltschka}, \citenamefont {Assig}, \citenamefont {Hardock},
  \citenamefont {Etzkorn}, \citenamefont {Ast},\ and\ \citenamefont
  {Kern}}]{supp_jack2015}%
  \BibitemOpen
  \bibfield  {author} {\bibinfo {author} {\bibfnamefont {B.}~\bibnamefont
  {Jaeck}}, \bibinfo {author} {\bibfnamefont {M.}~\bibnamefont {Eltschka}},
  \bibinfo {author} {\bibfnamefont {M.}~\bibnamefont {Assig}}, \bibinfo
  {author} {\bibfnamefont {A.}~\bibnamefont {Hardock}}, \bibinfo {author}
  {\bibfnamefont {M.}~\bibnamefont {Etzkorn}}, \bibinfo {author} {\bibfnamefont
  {C.~R.}\ \bibnamefont {Ast}}, \ and\ \bibinfo {author} {\bibfnamefont
  {K.}~\bibnamefont {Kern}},\ }\href {\doibase
  http://dx.doi.org/10.1063/1.4905322} {\bibfield  {journal} {\bibinfo
  {journal} {App. Phys. Lett.}\ }\textbf {\bibinfo {volume} {106}},\ \bibinfo
  {eid} {013109} (\bibinfo {year} {2015}),\
  http://dx.doi.org/10.1063/1.4905322}\BibitemShut {NoStop}%
\bibitem [{\citenamefont {Misra}\ \emph {et~al.}(2013)\citenamefont {Misra},
  \citenamefont {Zhou}, \citenamefont {Drozdov}, \citenamefont {Seo},
  \citenamefont {Urban}, \citenamefont {Gyenis}, \citenamefont {Kingsley},
  \citenamefont {Jones},\ and\ \citenamefont {Yazdani}}]{supp_misra2013}%
  \BibitemOpen
  \bibfield  {author} {\bibinfo {author} {\bibfnamefont {S.}~\bibnamefont
  {Misra}}, \bibinfo {author} {\bibfnamefont {B.~B.}\ \bibnamefont {Zhou}},
  \bibinfo {author} {\bibfnamefont {I.~K.}\ \bibnamefont {Drozdov}}, \bibinfo
  {author} {\bibfnamefont {J.}~\bibnamefont {Seo}}, \bibinfo {author}
  {\bibfnamefont {L.}~\bibnamefont {Urban}}, \bibinfo {author} {\bibfnamefont
  {A.}~\bibnamefont {Gyenis}}, \bibinfo {author} {\bibfnamefont {S.~C.~J.}\
  \bibnamefont {Kingsley}}, \bibinfo {author} {\bibfnamefont {H.}~\bibnamefont
  {Jones}}, \ and\ \bibinfo {author} {\bibfnamefont {A.}~\bibnamefont
  {Yazdani}},\ }\href {\doibase http://dx.doi.org/10.1063/1.4822271} {\bibfield
   {journal} {\bibinfo  {journal} {Rev. Sci. Inst.}\ }\textbf {\bibinfo
  {volume} {84}},\ \bibinfo {eid} {103903} (\bibinfo {year} {2013}),\
  http://dx.doi.org/10.1063/1.4822271}\BibitemShut {NoStop}%
\bibitem [{\citenamefont {Ambegaokar}\ and\ \citenamefont
  {Baratoff}(1963)}]{supp_ambegaokar1963}%
  \BibitemOpen
  \bibfield  {author} {\bibinfo {author} {\bibfnamefont {V.}~\bibnamefont
  {Ambegaokar}}\ and\ \bibinfo {author} {\bibfnamefont {A.}~\bibnamefont
  {Baratoff}},\ }\href {\doibase 10.1103/PhysRevLett.10.486} {\bibfield
  {journal} {\bibinfo  {journal} {Phys. Rev. Lett.}\ }\textbf {\bibinfo
  {volume} {10}},\ \bibinfo {pages} {486} (\bibinfo {year} {1963})}\BibitemShut
  {NoStop}%
\end{thebibliography}%

\end{document}